\documentclass[sigconf, nonacm]{acmart}

\usepackage{multirow}
\usepackage{color}
\usepackage{colortbl}
\usepackage{stfloats}
\usepackage{subcaption}
\usepackage{enumitem}
\usepackage{graphicx}
\usepackage{subcaption}
\usepackage{tabularx}
\usepackage{longfbox}
\usepackage{soul}
\usepackage[dvipsnames]{xcolor}
\usepackage{pifont}
\usepackage{calc}

\usepackage{kotex} 

\makeatletter
\newdimen\@tempdimd
\makeatother

\newfboxstyle{boxparam}{padding=1.5pt, padding-left=2pt, padding-right=2pt, height=7.5pt, background-color=lightgray, border-radius=2pt, border-right-width=1pt, border-bottom-width=1pt, border-color=gray}

\newcommand{\rectwrapsmall}[2]{\lfbox[boxparam, border-radius=0pt, padding-left=2pt, padding-right=2pt, height=5.5pt, border-width=0pt, background-color=#1]{\sffamily{\textcolor{white}{#2}}}}

\definecolor{quotebackground}{HTML}{EFEFEF}
\definecolor{tableheader}{HTML}{EFEFEF}
\definecolor{tablegrayline}{HTML}{e0e0e0}

\definecolor{desirable}{HTML}{31ac65}
\definecolor{undesirable}{HTML}{ef6c47}

\newcommand{\sysname}{\textsc{AutiHero}}
\newcommand{\creator}{Creator}
\newcommand{\reader}{Reader}

\newcommand{\interaction}{Relationship}
\newcommand{\norm}{Social Rules}
\newcommand{\adl}{Healthy Habits}

\newcommand{\eg}{\textit{e.g.}}
\newcommand{\ie}{\textit{i.e.}}
\newcommand{\cf}{\textit{c.f.}}

\newcommand{\reportstats}[5]{$SD=#1$; range: #2--#4}

\newcommand{\labelphantom}[1]{%
  \parbox{0pt}{\phantomsubcaption\label{#1}}%
}

\newboolean{clean}
\setboolean{clean}{false} 

\newcommand{\revised}[1]{\ifthenelse{\boolean{clean}}{#1}{\textcolor{blue}{#1}}}

\newboolean{cameraclean}
\setboolean{cameraclean}{true} 

\newcommand{\cameraready}[1]{\ifthenelse{\boolean{cameraclean}}{#1}{\textcolor{blue}{#1}}}

\newcommand{\deleted}[1]{%
  \ifthenelse{\boolean{clean}}{}{%
    \textcolor{cyan}{\st{#1}}%
  }%
}

\newcommand{\deletedsubsection}[1]{%
  \ifthenelse{\boolean{clean}}{}{%
    \subsection{\textcolor{cyan}{[Deleted] #1}}%
  }%
}

\newcommand{\circledigit}[1]{\textbf{\normalsize{\textsf{\textcircled{\footnotesize{#1}}}}}}

\newcommand{\ipstart}[1]{\vspace{1mm} \noindent{\textbf{\textit{#1.}}}}

\newcommand{\bpstart}[1]{\vspace{1mm} \noindent{\textbf{#1}}}

\makeatletter
\def\sectionautorefname{\S\@gobble}
\def\subsectionautorefname{\S\@gobble}
\def\subsubsectionautorefname{\S\@gobble}
\makeatother

\AtBeginDocument{%
  }

\setcopyright{none}
\copyrightyear{}
\acmYear{}
\acmDOI{}
\acmConference[UIST '26]{UIST}{Nov 02--05,
  2026}{Detroit, MI, USA}
\acmISBN{}

\begin{document}

%
\title{\sysname{}: Engaging Parents in Creating Personalized, Multi-path~Social Narratives for Autistic Children}





\settopmatter{authorsperrow=5}

\author{Jungeun Lee}
\authornote{Jungeun Lee conducted this work as a research intern at NAVER AI Lab.}
\orcid{0000-0002-2025-0870}
\affiliation{%
  \institution{POSTECH}
  \country{Republic of Korea}}
\email{jelee@postech.ac.kr}

\author{Kyungah Lee}
\orcid{0009-0009-7872-6454}
\affiliation{%
  \institution{Dodakim Child Development Center}
  \country{Republic of Korea}
}
\email{hiroo6900@hanmail.net}

\author{Inseok Hwang}
\orcid{0000-0001-7370-3944}
\affiliation{%
  \institution{POSTECH}
  \country{Republic of Korea}}
\email{i.hwang@postech.ac.kr}

\author{SoHyun Park}
\orcid{0000-0001-8703-0584}
\affiliation{%
  \institution{NAVER Cloud}
  \country{Republic of Korea}}
\email{sohyun@snu.ac.kr}

\author{Young-Ho Kim}
\orcid{0000-0002-2681-2774}
\affiliation{%
  \institution{NAVER AI Lab}
  \country{Republic of Korea}
}
\email{yghokim@younghokim.net}

\begin{abstract}
Social narratives help autistic children understand and navigate social situations through stories. To ensure effective practice, they often require significant time and effort from parents in customizing the narrative materials and delivering repeated instructions on them. We present \sysname{}, an generative AI-based social narrative system, which supports parents to create personalized, multi-path stories targeting specific behavior of their autistic children, while enabling them to explore behavioral choices and causal consequences together in reading. A two-week deployment study with 16 autistic child--parent dyads showed that parents actively created, adapted, and read stories with their children, with increased confidence in everyday behavioral guidance. Our work contributes to real-world-contextualized text+image content creation approaches harnessing generative AI, ensuring user-aligned application in sensitive contexts involving autistic children and their parents. 



\end{abstract}





\begin{teaserfigure}
 \centering
 \includegraphics[width=\textwidth]{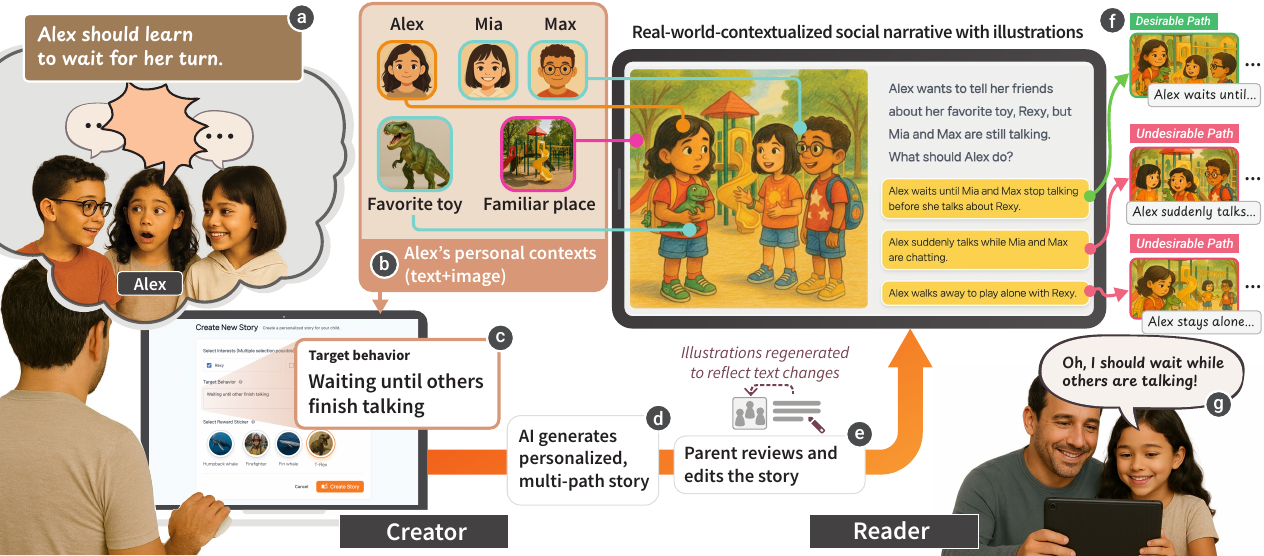}
 \caption{\sysname{} helps parents create personalized stories to guide their children's behaviors, leveraging generative AI. Parents first register their child's familiar contexts (\circledigit{b}) such as friends, belongings, and places, into the \creator{} app using both text and images. Then they enter a specific target behavior for their child they want to address in daily life (\circledigit{a})---\eg{}, waiting until others finish talking---into the system (\circledigit{c}). The system then generates a story (\circledigit{d}) with characters and environments grounded by the child's real-world contexts, and parents can review and edit the story (\circledigit{e}). The narrative demonstrates multiple behavioral choices (\circledigit{f}) with different outcomes. Using the \reader{} app, parents and the child can read the story together (\circledigit{g}).}
 \label{fig:teaser}
 \Description{Figure 1 illustrates the overall workflow of AutiHero, a generative AI–based system that helps parents create personalized stories to support children’s behavioral development.
Left side (Creator app):
A parent identifies a target behavior (e.g., waiting until others finish talking) and inputs it into the Creator interface along with the child’s interests and context items.
Example shown: Alex's dad types the behavior ``waiting until others finish talking.''
Right side (Reader app):
The system generates a story in which Alex, the child, is the protagonist.
The story integrates Alex's real friends, favorite toy (Rexy), and frequently visited playground into the narrative context.
The generated story includes decision-making prompts for the child, such as whether Alex should wait until her friends finish talking before speaking.
Parents and the child read the story together on the Reader app, reinforcing the target behavior.
The figure combines illustrated story scenes, photos of real-world context items, and app screenshots to show how personalization bridges children’s daily life and the story content.}
\end{teaserfigure}


\maketitle

\section{Introduction}
Autistic children\footnote{In this work, we use identity-first language (\eg{}, autistic children) rather than person-first language (\eg{}, child with autism), considering the preferences of autistic individuals~\cite{Lorcan2016} and recent academic trends~\cite{Brian2002}.} often experience difficulty understanding social situations due to challenges in interpreting implicit cues and inferring others' intentions~\cite{callenmark2014explicit, keifer2020prediction, pelzl2023reduced, jacob2012cerebral, argyle1970communication}.
To support social understanding, behavioral therapists often use \textit{social narratives}~\cite{leaf2020critical}---structured stories that depict specific social situations---to guide expected behaviors. 
These narratives provide concrete explanations of social expectations and offer parents a structured way to discuss and rehearse everyday situations with their children~\cite{zimmerman2017beyond}.

However, using social narratives in everyday parenting requires parents to continually adapt stories to new situations their child encounters, integrating them into daily routines~\cite{barry2004using, karwal2007use} and tailoring them to the child's circumstances~\cite{edwards2021personalization}. 
In addition, crafting social narratives demands strict requirements rooted in autistic children's cognitive and perceptual characteristics~\cite{gray1993social,ulu2024power}. Experts emphasize that situations in the story should be grounded in familiar, everyday contexts and conveyed using literal, non-metaphorical language~\cite{vicente2023accounting, lampri2024figurative}. Also, the story needs to be decomposed into step-by-step progressions with minimal contextual noise, reducing ambiguity while preserving essential cues. 
Furthermore, in everyday use, social narratives need to support opportunities for rehearsal, allowing children to simulate different responses within the same situation to facilitate transfer to real-world behavior~\cite{barry2004using}. 
This compounded demand imposes substantial burden on parents~\cite{trembath2019systematic}.

Recently, generative AI (GenAI) technologies such as large language models (LLMs) and text-to-image models demonstrate generating personalized stories on demand (\eg,~\cite{geministorybook}, \cite{lee2024open}). However, applying them to narrative construction and image generation is still limited at best. Stories, especially for autistic audience, require a higher degree of explicitness and stability: causality between actions and outcomes must be clearly articulated~\cite{landau2025context}, and visual representations of characters, objects, and environments must remain consistent across scenes~\cite{muller2008children}. In practice, GenAIs often fail to maintain such causal structure~\cite{wu2025llm} and cross-scene consistency~\cite{shin2025generating}, at times producing inconsistent and even inappropriate narratives.

To address barriers of social narrative practice at home, we present \sysname{} (\autoref{fig:teaser}), a GenAI-based system for parents to create personalized, multi-path social narratives and read them together with their children. Based on the parent-specified target behavior, child's interest (\eg{}, a favorite toy), and personal contexts provided in text and image, \sysname{} generates a social narrative with multiple pages and illustrations, which then allows parents to review, edit, or regenerate. 
The design of \sysname{} was informed by formative interviews (\autoref{sec:formative}) with autism experts ($N=10$). 
To ensure that the generated stories foster exploration and comparison of alternative behaviors through clearly structured step-by-step progression, the system applies the following principles across multiple levels of narrative construction:
(1) A \textit{story} consists of explicit \textbf{desirable and undesirable paths} within the context of the target behavior, ensuring that differences in outcomes can be directly attributed to differences in actions.
(2) The system expands each \textit{path} into a \textbf{semantically defined steps} (\eg{}, challenge, decision, and repair in \autoref{fig:system_storystructure}), preventing implicit jumps and making behavioral causality observable. 
(3) Each \textit{page}, corresponding to a step, contains text and an illustration grounded in the child's real-life context. The system ensures \textbf{consistency of context-specific elements across pages} and alignment between text and images through context-grounded generation. 

We conducted a two-week deployment study (\autoref{sec:deployment}) with 16 parent--autistic child dyads to explore how parents use \sysname{} for creating and reading the narratives with their children, and to understand how \sysname{} supports parents' confidence and practices in everyday behavioral guidance and children's social behavior.
%
A total of 218 stories was created, and an average of 4.25 stories was read per day.
Quantitative and qualitative findings show that parents utilized \sysname{} to explain various social situations and found the stories a useful means to respond to challenging moments.

The key contributions of this work are as follows:

\begin{enumerate}[leftmargin=*, itemsep=2pt, topsep=2pt]
\item The design and implementation of \sysname{}, a GenAI-based system that supports parents to create personalized social narratives for their autistic children, using a multi-path narrative structure and a context-grounded text/image generation pipeline, and post-hoc review and editing.
\item Empirical findings from a two-week deployment study involving 16 parent–child dyads, demonstrating how AI-generated social narratives can support parents in responding to everyday situations with their autistic children.
\item Implications for designing GenAI-based social narrative technologies that support parental guidance and reflection for fostering effective social communication for autistic children.
\end{enumerate}

\section{Related Work}
In this section, we cover related work regarding social narrative methods and challenges of adapting them. We then discuss personalization and interactivity of storytelling systems for children.

\subsection{Social Narratives for Autistic Children}

Autistic children often struggle with social interactions due to differences in processing communication and interpreting social cues~\cite{jellema2009involuntary, forby2023reading}. 
Their literal style of thinking~\cite{happe1997central, jolliffe1999test} can make it hard for them to understand facial expressions, gestures, tone, or unwritten rules~\cite{frith2003autism}, motivating a range of intervention strategies. 

One common approach is social narratives, which depict desirable behaviors through stories and help autistic children respond appropriately in social situations. These include Social Stories~\cite{gray1993social}, power cards~\cite{gagnon2001power}, Comic Strip Conversations~\cite{gray1994comic}, and cartooning~\cite{coogle2018social}.
Social Stories describe everyday situations in narratives 
to clarify what happens, what behavior is expected, and why~\cite{gray1993social}.
Power cards employ visual prompts linked to a child's special interests 
to illustrate desirable behaviors in an engaging format. 
Another widely used method is video modeling~\cite{delano2007video}, which presents video demonstrations of target behaviors---\eg{}, greeting a peer---so that children can learn by repeatedly watching these behaviors in action. 

Despite clinical evidence (\eg{},~\cite{o2015relative, gandhi2024comparative, charlop2000comparison, thiemann2001social, brownell2002musically}) on their effectiveness, practical barriers remain.
Crafting compelling social narratives, even in text-only form, is not straightforward. For example, Social Stories impose ten strict rules~\cite{gray1993social} on content and format that can be difficult for less experienced individuals to follow. While video modeling is known to be more effective~\cite{o2015relative, gandhi2024comparative, charlop2000comparison}, it requires video production skills and efforts. This poses barriers when parents apply these approaches at home.

%

Furthermore, prior work suggests that effective use of social narratives requires continuous adaptation to children's specific contexts and everyday situations~\cite{wright2012utilizing}. To reduce authoring burden, GenAI has been leveraged to develop online Social Story generators (\eg,~\cite{writifyai, ella, socialstorycreator}) for automatic generation. While these tools follow established Social Story guidelines~\cite{gray1993social}, they provide limited support for rehearsal and behavioral exploration, which prior work highlights as essential for real-world behavior transfer~\cite{barry2004using}.
Because parents have detailed knowledge of their children's interests and contexts, they are well positioned to tailor guidance to individual needs. Therefore, there remains a need for tools that better support parent-driven personalization of social narratives in everyday contexts.
Our work expands the line of research on technology-supported interventions to foster social and cognitive development in autistic children~(\eg,~\cite{bei2024starrescue, gagan2023preparing, park2025lessons, li2021faceme}). Unlike prior work focused mainly on instructional training or creating immersive learning environments in VR/AR, we support parent-driven story creation and adaptation based on social narrative approaches.

\subsection{Storytelling Systems for Children}
Research on storytelling systems for children has explored ways to adapt narratives to individual needs. A large body of work focused on personalization, tailoring stories to reflect children’s characteristics, interests, and environments. For example, People in Books~\cite{follmer2012people} incorporates familiar people into storybooks, and self-avatar representations have been shown to enhance immersion and engagement~\cite{zarei2020investigating}. Personalized digital books have also demonstrated benefits for language development~\cite{kucirkova2021empirical, kucirkova2014reading, lee2024open} and verbal expression~\cite{kucirkova2013parents}. These approaches highlight how aligning stories with children's lived experiences can increase engagement and relevance. In these works, however, personalization has primarily focused on adapting the story content, with limited support for guiding how such narratives are used to support learning or reflection in everyday contexts.

Prior work has also explored interactive storytelling, where users can influence story progression through multiple paths. Branching or interactive narratives have been used to support engagement and understanding by enabling users to navigate nonlinear story structures, make choices and exercise agency~\cite{roth2025stories}, and explore alternative outcomes~\cite{kim2011programming}. 
Recent systems extend this paradigm using large language models to dynamically adapt narratives to learners' inputs~\cite{cheng2025oak}.
However, little work has examined their use for children. Even fewer efforts have considered how to design multi-path structures for behavioral understanding, such as helping children reason about everyday actions and their consequences.

In light of social narrative approaches, we aim to extend earlier personalization and interactive storytelling approaches for autistic children. \sysname{} supports parents to create personalized social narratives that are grounded in children's everyday contexts, as well as offers within each narrative structured decision points to explore different behavioral possibilities and their outcomes, allowing the children to observe how actions lead to social consequences.

\section{Formative Study}\label{sec:formative}
To inform the design of \sysname{}, we conducted formative interviews with ten professionals (E1--10; see \autoref{tab:formative_demographics} in Appendix) specializing in autism support and social communication development and who have hands-on experiences working with and/or as parents of autistic children.
The experts were from various professional settings, spanning from education to psychotherapy. 

We conducted 1-hour interviews with each expert, covering
(1) the real-world challenges parents face in teaching social communication to autistic children; (2) the experts' hands-on experiences in implementing and customizing social narratives. To explore their perspectives around GenAI in this context, we presented a scenario video of AI-based story creation as a probe.
We offered a 100,000 KRW (approx. 73 USD) gift card as compensation.

The interviews were audio-recorded, anonymized and transcribed. We analyzed the transcripts using thematic method~\cite{braun2006using}, 
deriving key findings that informed the design goals of \sysname{}.


\subsection{Finding 1: Barriers to Parent-Led Social Narrative Implementation}\phantomsection{}\label{sec:formative_finding1}
Although social narratives are originally designed to support both parents and educators in facilitating everyday learning at home~\cite{gray2000new}, experts noted that many parents struggle to implement them consistently. Many experts (E1, E3-4, E6, E8-9) pointed out that caregiving demands often leave parents physically and emotionally exhausted, limiting their capacity to engage with additional intervention strategies. E3 also emphasized, \textit{``Parents' self-efficacy plays a critical role in whether such tools are adopted in practice. In other words, parents need to feel confident that they can effectively teach their own child.''}

\bpstart{Design Goal 1 (DG1):} Engage parents in the authoring process by supporting parent-specified topics and contexts, low-burden story generation, and iterative reviewing and editing workflows.

\subsection{Finding 2: Challenges in Creating Contextualized Social Narratives}
Experts consistently emphasized that social narratives for autistic children must be carefully tailored to their interests, linguistic abilities, and everyday environments. 
This is because they often rely on explicit and concrete cues to understand social situations, and may have difficulty interpreting implicit meanings. 
This requires incorporating specific details from the child's daily routines. To achieve this, experts gather rich contextual information and iteratively refine materials with parents. E1 described, \textit{``Even if we include certain images, parents might point out something missing. One parent said, `My child always touches a specific signboard before crossing the street. That signboard has to appear in the picture.' These small details often make a real difference.''} However, creating and refining such personalized materials requires a significant effort. 
E6 remarked, \textit{``Sometimes we spend more time learning Photoshop than playing with the child. I also tried generating images with AI, but it's hard to adapt them to match our emotional and cultural setting.''}

\bpstart{Design Goal 2 (DG2):}
Set the child as the protagonist to ground the narrative in the child's own perspective, enabling behavioral causality to become directly observable and interpretable.

\bpstart{Design Goal 3 (DG3):} 
Reflect the child's everyday environment, including personal interests, people, and places, in the narratives to ensure consistency with the child's lived experiences.

\subsection{Finding 3: Need for Simulation and Practice}
Experts highlighted the necessity of rehearsal and role play for social narratives, because autistic children often fail to transfer what they have learned into actual behavior due to the lack of opportunities for simulation and practice. They also stressed the value of repeatedly presenting children with challenging situations, teaching them how others might feel, and guiding them toward appropriate responses. E8 explained, \textit{``{We break down situations into specific cases and discuss how people typically respond in those situations, as well as what more appropriate response would look like.}''}


\bpstart{Design Goal 4 (DG4):} 
Support behavioral rehearsal by presenting contrasting behavioral choices and their consequences, enabling children to explore and compare alternative actions.

\section{\sysname}

\sysname{} is designed to support parents to create personalized, multi-path social narratives for autistic children in guiding social behavior (DG1). 
In this section, we describe the key approaches for story design and generation, as well as the implementation details.

\subsection{Key Approaches}

To ensure reliable creation of social narratives that reflect the child's real-life context while complying with the design goals, \sysname{} follows a multi-level construction approach. 

\subsubsection{Story-level: Enabling behavioral comparison through structured multi-path narratives}

To support behavioral comparison to help autistic children better understand the reasoning behind social behaviors in a story (DG4), \sysname{} presents alternative actions and their consequences. As conventional social narrative methods usually offer only general design guidelines~\cite{gray1993social}, experts often manually write the stories themselves. To scaffold the narrative construction,
\sysname{} incorporates a multi-path story structure that organizes alternative behavioral paths around a moment of decision for the target behavior, where the child must choose between multiple actions. The paths are either desirable or undesirable, and their composition and progression vary by behavior type (\autoref{sec:story:path_by_topic_type}), allowing the system to reflect different social dynamics and common challenges while enabling clear comparison across alternatives.

\subsubsection{Path-level: Structuring causally-connected narrative sequence}

\sysname{} ensures that each path explicitly conveys how the child's actions lead to outcomes, while placing the child at the center of the narrative (DG2). 
To achieve this, \sysname{} structures each path as a sequence of semantically defined steps (\eg{}, challenge, decision, consequence, repair; \autoref{fig:system_storystructure}), linking the child's actions to subsequent social responses and outcomes. Desirable paths emphasize positive social responses to appropriate behavior, whereas undesirable paths emphasize negative social consequences of inappropriate behavior. Notably, undesirable paths always demonstrate how to resolve the worsened situation, to help children understand not only what went wrong but also how to repair the situation appropriately. 
The path-based story structure is described in \autoref{sec:system_story_structure}.


\subsubsection{Scene-level: Maintaining consistent and real-life-contextualized entity representations}
Each story page contains text and an illustration depicting a scene.
Based on expert feedback that photorealistic details may distract autistic children from the key messages and that their perfectionistic tendencies may increase anxiety due to subtle mismatches with real-world objects~\cite{yang2026autiverse}, \sysname{} adopts a colored-pencil illustration style, while preserving and translating salient characteristics (\eg, hairstyle, accessories, outfit) of parent-provided real-life context (see \autoref{fig:teaser}).
To capture key aspects of the situation---including the child's actions, surrounding context, and relevant social cues (DG3)---the system generates each scene illustration by enforcing entity consistency across scenes and text--image alignment (\autoref{sec:system:pipeline:illustration}).
It first produces a structured scene description specifying the action, participating entities, and environmental settings, and maintains shared representations of entities to ensure consistent appearances. These descriptions guide image generation, aligning visuals with the narrative while preserving coherence within each scene. This process also reduces the parental burden of manually specifying detailed visual and contextual elements (DG1).

\subsection{Multi-Path Story Structure}
\label{sec:system_story_structure}
\begin{figure*}
    \centering
    \includegraphics[width=\textwidth]{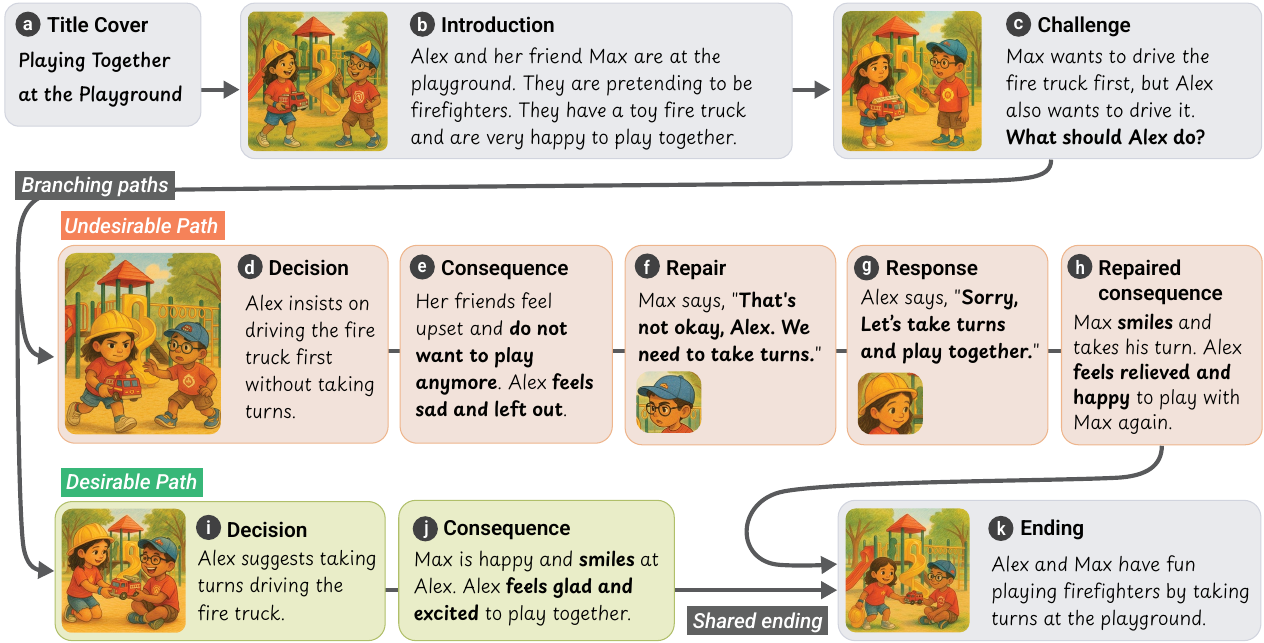}
    \caption{The multi-path structure of the \sysname{} stories. In the \textbf{Challenge} page (\circledigit{c}) the child is asked to select one of the given paths (\circledigit{d} or \circledigit{i}). The undesirable paths always have a turning point, \textbf{Repair} (\circledigit{f}), which addresses the negative \textbf{Consequence} (\circledigit{e}) and eventually leads the story to the positive \textbf{Ending} (\circledigit{k}) same as the desirable path.}
\label{fig:system_storystructure}
\Description{Figure 2 illustrates an example story structure with desirable and undesirable paths. The purpose of this figure is to demonstrate how children can navigate social situations by making choices and understanding the consequences of their actions.
The structure is organized as a flowchart with illustrated panels and accompanying text.
1. The story begins with the Cover (a), introducing Alex and her friend Max playing firefighters with a toy fire truck.
2. The introduction (b) describes the playground setting and their pretend play. 
3. The challenge (c) poses a question: both Alex and Max want to drive the fire truck first---what should Alex do?
From here, two main paths branch out:
4-1. Desirable path (green branch):
Desicion (d): Alex suggests taking turns.
Consequence (e): Max smiles, and both children feel glad and excited to play together.
4-2. Undesirable path (orange branch):
Decision (f): Alex insists on driving first without taking turns.
Consequence (g): Friends feel upset and do not want to play anymore; Alex feels left out.
Repair (h): Max explains that taking turns is important.
Response (i): Alex apologizes and suggests playing together.
Repaired Consequence (j): Max smiles, and Alex feels relieved and happy again.
Finally, both paths converge at the Ending (k), where Alex and Max enjoy playing firefighters by taking turns.}
\end{figure*}


Four authors (three HCI researchers, one autism expert) collaborated to design and refine the story structure.
We reviewed guidelines for social narratives and example Social Stories and power cards. 
We then conducted brainstorming sessions to structure concise but meaningful story pages, given that Social Stories are recommended to have no more than 12 sentences~\cite{gray1993social} for children.

\subsubsection{Story Paths and Flows} 

\autoref{fig:system_storystructure} illustrates the finalized story structure. A story begins with \textbf{Title} (\circledigit{a}) and \textbf{Introduction} (\circledigit{b}) that establish the characters and context, followed by \textbf{Challenge} (\circledigit{c}) where the child chooses between alternative \textbf{Decision}s (\circledigit{d} and \circledigit{i}). 
Each story includes one \rectwrapsmall{desirable}{desirable} path and one or two \rectwrapsmall{undesirable}{undesirable} paths, which diverge in their \textbf{Consequences} (\circledigit{e} and \circledigit{j}) but converge to a shared positive \textbf{Ending} (\circledigit{k}).
The undesirable paths include a \textbf{Repair} (\circledigit{f}) sequence, where another character provides suggestions that can resolve the situation. In \textbf{Response} (\circledigit{g}), the child responds by taking appropriate actions. 
A \textbf{Repaired Consequence} (\circledigit{h}) then presents the positive outcomes resulting from this repair, and transitions to the positive \textbf{Ending} (\circledigit{k}).

Theory of mind---inferring the others' mental states---is a common difficulty that autistic children face~\cite{happe1997central, baron1985tom}. To help comprehend of others' emotions, the story explicitly presents characters' emotions alongside their observable responses---\eg, ``Max (friend) is \textit{happy} (emotion) and \textit{smiles} (response)'' in Consequence (\circledigit{j}).

\subsubsection{Story Topics}\label{sec:story:path_by_topic_type}
To construct narratives appropriately within different situations, we defined three topic types---\textit{\interaction{}}, \textit{\norm{}}, and \textit{\adl{}}---each with a distinct flow and composition of desirable and undesirable paths. These topics were informed by social narrative literature (\eg,~\cite{como2024scoping}) and our formative study. 
The \textbf{\interaction{}} type covers situations requiring social communication or cooperation, such as \textit{asking a friend for permission to try their toy} or \textit{asking a teacher for help when difficulties arise}. Its undesirable paths include behaviors breaking rules (\eg{}, \textit{taking away friend's toy}) and avoiding social engagement (\eg{}, \textit{leaving to play alone}). 
The \textbf{\norm{}} type involves behaviors governed by contextual or cultural expectations where violations are socially discouraged and often corrected by others---\eg{}, \textit{keeping calm during prayer at church}. 
The \textbf{\adl{}} type concerns daily routines that support independent and sustainable lifestyle, such as \textit{washing hands before meals}, where parents play a central role in shaping children’s daily habits and guiding the acquisition of self-care skills.

\subsection{Generative Pipelines}
\begin{figure*}[t]
    \centering
    \includegraphics[width=\textwidth]{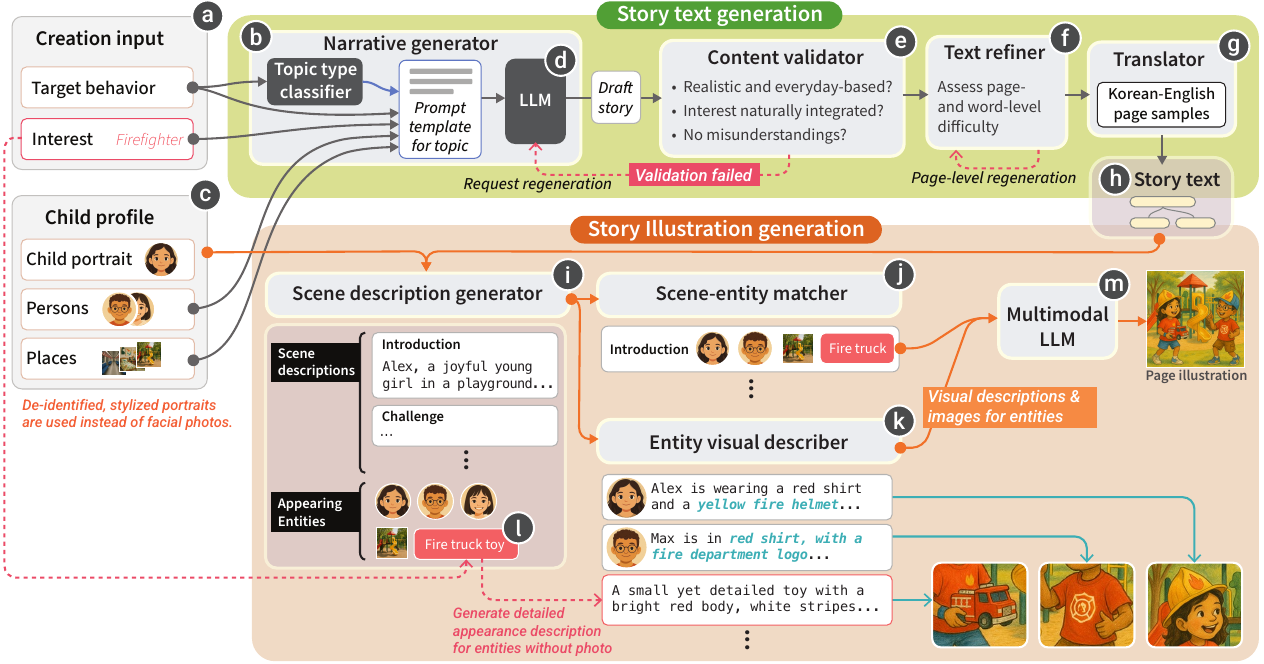}
    \caption{Pipelines for generating story text and illustrations. The parent enters the target behavior and selects an interest to be featured in the story (\circledigit{a}).
    The Narrative generator (\circledigit{b}) produces a draft story, which is then validated and refined by the Content validator (\circledigit{e}) and Text refiner (\circledigit{f}). Next, the Scene description generator (\circledigit{i}) creates scene descriptions for each page, taking the child profile (\circledigit{c}) into account. Finally, with support from the Scene--entity matcher (\circledigit{j}) and Entity visual describer (\circledigit{k}), a multimodal LLM (\circledigit{m}) generates illustrations for each page that remain consistent with the story context.}
    \label{fig:system_pipeline}
    \Description{Figure 4 presents the pipeline for generating personalized story text and corresponding visual illustrations. The process integrates creation input (target behavior, child interests, and child profile information) with multiple text and image generation components.

1. Story text generation (a–h):
The topic type classifier (b) categorizes the input into relationship, social rules, or healthy habits.
The story generator (d) produces an initial draft, which is checked by the content validator (e) for realism, integration of interests, and clarity.
If validated, the draft is further refined by the text refiner (f) for difficulty level and, when needed, translated by the story translator (g). The output is the final story text (h).
2. Preprocessing for illustration generation (i–k):
The scene description generator (i) creates descriptions of each story section.
The entity matcher (j) links textual references (e.g., fire truck, child’s name) to user-provided images and entities.
The entity describer (k) produces detailed visual descriptions of characters and objects (e.g., clothing, colors, logos).
3. Illustration generation (m):
The image generator receives prompts derived from scene and entity descriptions, then produces illustrations aligned with the story context. Images are generated in parallel for all story sections to ensure consistency.
The figure uses arrows and labeled modules to highlight the sequential and interconnected flow from text generation to illustration synthesis.}
\end{figure*}

\autoref{fig:system_pipeline} illustrates the generative pipelines of \sysname{} for generating a story and associated illustrations. 

\subsubsection{Story and Path Generation}
When a parent initiates a new story~(\autoref{fig:system_pipeline}-\circledigit{a}), the \textbf{Narrative generator} (\circledigit{b}) first produces a draft multi-path story based on the target behavior, selected interests, and entities in the child profile (\circledigit{c}), including persons and places. The LLM (\circledigit{d}) generates the story using a prompt instantiated from a template corresponding to the topic type derived from the target behavior (\autoref{fig:story_generation_prompt} in Appendix). The prompt instructs to select appropriate persons and places to include in the story.

The story undergoes multiple refinement steps. First, the \textbf{Content validator} (\circledigit{e}) assesses the quality of the story \textit{content}, checking whether it supports clear and interpretable communication for autistic children (\eg, avoiding metaphorical representations) and whether it is well grounded in the child's real-world context, particularly the interest. The story is iteratively regenerated until it satisfies all criteria.
Next, the \textbf{Text refiner} (\circledigit{f}) adjusts the text to match the reading level of elementary school children. Each page's text is regenerated in simpler phrasing if deemed difficult for U.S. Grade 5 readers, based on Flesch-Kincaid Grade Level~\cite{flesch1948new}, or if it contains any words above B2 (Upper Intermediate) according to the Common European Framework of Reference for Languages~\cite{council2001common}.

Finally, the \textbf{Story translator} (\circledigit{g}) translates the refined English story into Korean. As automated readability assessment for Korean remains limited, we did not apply a validation step to translated outputs. Instead, we constructed 190 manually refined English--Korean page pairs from 15 generated stories. The translator LLM uses the 22 pages with the highest vector similarity to the input story as few-shot samples.

For the story generation pipeline, we employed OpenAI's \texttt{gpt-4o} for classification and generation, while \texttt{gpt-4.1} handled the translation. The entire story text generation typically takes 1--2 minutes.

\subsubsection{Scene Illustration Generation}\label{sec:system:pipeline:illustration}
To ensure coherence of content and contextual entities across pages, the \textbf{Scene description generator} (\autoref{fig:system_pipeline}-\circledigit{i}) first produces scene descriptions for all pages, based on the full story text (\circledigit{h}), along with a list of entities (persons, objects, and places). The \textbf{Scene--Entity matcher} (\circledigit{j}) then assigns these entities to each scene based on both the scene descriptions and the overall story context.
This step ensures the inclusion of entities that are not explicitly mentioned in the text but are implicitly expected to appear in the scene (\eg{}, characters from previous scene), preserving continuity across pages. 

In parallel, the \textbf{Entity visual describer} (\circledigit{k}) generates visual descriptions for each entity; for entities registered in the system (\ie{}, interests and entities in the child profile \circledigit{c}), associated images are used as inputs. For persons, additional appearance details are generated based on the story context, ensuring that attributes such as outfits reflect the narrative setting (\eg{}, wearing a swimsuit in at the beach). For entities either without images or not in the profile, the system produces detailed descriptions to guide their visual representation (\eg{}, `\textit{fire truck toy}'--\circledigit{l}).
The preprocessed descriptions are cached and reused when a parent requests regeneration unless the page text has changed and no longer aligns with them.

Lastly, for each page, a \textbf{multimodal LLM} (\circledigit{m}) generates an illustration using the page's scene description and the associated entities' visual descriptions and images as multimodal input.

\sloppy{We used OpenAI's \texttt{gpt-4o} for description generation and entity matching, and \texttt{gpt-image-1} for image generation. Creating all illustrations for a story typically took 2 minutes with parallel generation of pages.}

\subsubsection{Privacy Safeguards}
Grounding illustrations on real-life contexts requires visual information, including facial images of the child and familiar people.
To safeguard personal information in model inputs, we used OpenAI Enterprise\footnote{https://openai.com/enterprise-privacy/}, which neither uses our input text and images for training nor retains them. To further minimize privacy risk associated with facial images, our backend server converts the parent-provided images of children and persons into stylized portraits (see \autoref{fig:system_pipeline}--\circledigit{c}), using \texttt{gpt-image-1}, removing identifiable facial details while preserving general visual characteristics. The original facial images are deleted after stylization, and only the resulting images are used as inputs.

\subsection{\sysname{} Apps: \creator{} and \reader{}}

\sysname{} consists of \textbf{\creator{}} (\autoref{fig:teaser}, left) for story creation on web and \textbf{\reader{}} (\autoref{fig:teaser}, right) for story reading on tablets. \creator{} allows parents to manage the child's real-life contexts (interests and the child profile---\circledigit{c} in \autoref{fig:system_pipeline}), and to create and review stories. 
Parents and children can read the created stories in \reader{} together. We chose to support reading on tablet to cultivate \textit{joint attention} of autistic children~\cite{bruinsma2004joint, leekam2000attention} and shared reading~\cite{lee2024open}. 

Here we present a usage scenario featuring David and his daughter Alex: \textit{David has an 8-year-old daughter, Alex, who is on the autism spectrum and considered high-functioning, being able to understand storybooks for kids. Alex often struggles with \textbf{taking turns during playtime with her friends}, which frequently leads to arguments. Wanting to help her understand this norm, David tries \sysname{} to create a personalized story that can gently guide Alex.} (See \autoref{fig:system_usageflow} in Appendix and supplemental video for detailed interactions.)

\subsubsection{Profile Management}



David enters \creator{} on web. For first-time use, he registers a photo of Alex, which the system converts into an illustrated portrait used by \creator{} to generate illustrations of a protagonist resembling Alex. He then adds \textbf{people} who regularly interact with Alex, including her family members and her close friends Mia and Max. He also registers familiar \textbf{places}---Alex’s bedroom, the playground, and the subway. He uploads photos and short descriptions for each. The interface also prompts him to provide Alex's \textbf{interests}. Alex plays with her dinosaur toy Rexy every day and dreams of becoming a firefighter. David registers Rexy and ``firefighter,'' adding simple descriptions and a photo for Rexy.


\subsubsection{Story Creation}
David starts creating a story by entering the required information. He first selects \textit{Firefighter} from Alex's registered interests and enters \textit{Taking turns during playtime} as the target behavior.
David then clicks the Create Story button. In three minutes, the system generates a story titled \textit{``Playing together at the Playground,''} which incorporates the firefighter interest through a firefighter-themed play scenario (see \autoref{fig:system_storystructure}). The story is set in Alex’s familiar playground and features Alex and her friend Max.

David reviews the generated story for quality. On one page, he notices an inconsistency in Max's clothing pattern and regenerates the illustration.
He also edits one of Max's dialogue lines to make it more natural and expressive.

\subsubsection{Story Reading}
In the evening, David reads the story with Alex to help her learn to take turns during playtime. Sitting together, he opens the \reader{} app on a tablet, selects the story, and begins reading it aloud.
Alex shows strong interest in the story, especially as she is the main character. She is also fascinated to see Max in the illustrations, resembling her real friend, along with the familiar playground. When they reach the Challenge page, David explains the situation and asks Alex what she would do. After some thought, Alex chooses \textit{``Alex insists on driving the fire truck first without taking turns''}. The story then shows Max becoming upset, while Alex feels frustrated and sad as the game does not go as planned. As Alex continues reading through the next pages, she learns how to apologize and resolve the conflict. 

David gently encourages Alex to try again. As they reread the story, Alex now selects the desirable path: \textit{``Alex suggests taking turns driving the fire truck.''} This time, the story shows Max smiling as he enjoys playing with Alex. At the end, 
David praises her for completing the story.

\subsection{Implementation}
We implemented the core system in Python running on a FastAPI~\cite{FastAPI} server that provides REST APIs for both \creator{} and \reader{} apps. The generative pipelines run on top of the LangChain~\cite{LangChain} framework to run the underlying LLM inference and image generation. The generated stories and user data are stored in a PostgreSQL~\cite{PostgreSQL} database on the server.
We built \creator{} as a web application using React.js~\cite{React}, and the \reader{} app using React Native~\cite{ReactNative} as a cross-platform tablet application running on both iPad and Android tablets. Both apps were written in TypeScript~\cite{TypeScript} and communicate with the server via REST API. 

\section{Deployment Study}\label{sec:deployment}
We conducted a two-week field deployment with 16 dyads of autistic children and their parents. We aimed to examine how parents use \sysname{} to create stories and read them with their children, and how this experience influenced how they approached everyday behavioral guidance and how their children responded to it. Our IRB approved the study protocol.

\subsection{Participants}
We recruited study participants from online communities of parents with autistic children. Our inclusion criteria for child participants were: (1) a diagnosis of Autism Spectrum Disorder classified as Level 1 or 2 autism per CDC guideline; (2) the ability to read and comprehend short stories with images and limited text, as typically targeted toward preschool and early elementary school children. One author who is a licensed counselor and authorized to diagnose autism, initially screened the respondents based on the description of the child's literacy and cognitive functioning.
%
%

The demographic information of the 16 parent (P)--child (C) dyads, including the children's autism level and literacy, is provided in \autoref{tab:deployment:demographic} in Appendix. Child participants (C1--16; two girls) were aged 7 to 12 ($M=8.56$). Twelve children were assessed as having Level 1 autism (formerly referred to as high-functioning autism). While the remaining four were assessed as having Level 2 autism in terms of verbal communication, they were also included in the study as they were considered to have potentially sufficient comprehension of images and text that \sysname{} offers. 
Each dyad was compensated 200,000 KRW (approx. 144 USD) as a gift card after the study.


\subsection{Procedure}
The field deployment consisted of three phases: (1) introductory session, (2) deployment, and (3) debriefing.
To each household, we deployed a Samsung Galaxy Tab S6 Lite Android tablet, which has a 10.4-inch (263mm) display with a 2000 $\times$ 1200 resolution (224 ppi). 
%


\ipstart{Introductory Session}
One researcher visited participants' homes and set up the tablet. 
Upon explaining the goal of the study and the protocol, we also provided a detailed explanation of how uploaded images would be handled and discarded, complying with the IRB guideline. All parents provided informed consent to participate.
We then provided a tutorial on how to create stories in \creator{} and read them in \reader{}.
The session took about ~45 minutes. 

\ipstart{Deployment} Participants freely used \sysname{} for a 14-day deployment period.
At 10 PM each evening, we sent parents a link to a survey asking about their daily experiences with \sysname{} with both multiple choice and open-ended questions. The survey asked parents to rate their children's engagement to the day's reading activity 
on a 5-point Likert scale and report any noteworthy comments regarding the day's story creation and reading activities.


\ipstart{Debriefing}
After deployment, we had a one-hour debriefing session with each participant. 
They first completed an exit survey with 7-point Likert scale questions across six subscales of the Technology Acceptance Model (TAM)~\cite{Venkatesh2008TAM3}
for both \creator{} and \reader{} apps, assessed separately.
The survey also asked parents to rate their perception of changes in their child's behavior, if any, and the corresponding shifts in parental response strategies, for each target behavior they had entered into the system, in a 5-point Likert scale.
%
Then we conducted a semi-structured interview. We asked about their experience of creating and reading stories with \sysname{}, their child's reactions to created stories and engagement in the reading activities, and the system's impact on parenting. 
%
All surveys and interviews were conducted with parents, as our system primarily targets them and directly collecting reliable responses from autistic children can be challenging, due to their communication and developmental constraints. Prior work on autistic children also notes these challenges and often relies on parent-reported data~\cite{choi2025aacesstalk, meerson2024provia}.

The debriefing interviews were audio-recorded, anonymized and transcribed. We conducted a thematic analysis~\cite{braun2006using}, whereby the first author generated initial code themes on Miro~\cite{miro}, which were then iteratively discussed and refined with two other researchers until consensus was reached on the final set of themes.

\section{Results}
We analyzed system logs, daily surveys, and the debriefing transcripts. We derived usage metrics from system logs and used daily surveys to capture children's engagement and parents' day-to-day experiences with story creation and reading. In this section, we report findings from the deployment study, including parents' story creation activities and their story reading. We then present parents' reflections on using \sysname{} for behavioral guidance.

\subsection{Story Creation}

We examine how parents engaged in creating stories with \sysname{}, \eg{}, overall creation patterns, types of stories and target behaviors, reactions to the creation process, and personalization.

\subsubsection{Creation and Editing Activities}
Over the two weeks, parents actively created stories with \sysname{}---a total of 218 stories, \ie{}, an average of 13.63 stories per parent (\reportstats{5.78}{6}{P3}{26}{P14}). 
Parents reported that they created stories whenever they thought of behaviors they wanted to teach. 

Parents also engaged in editing: Titles or page texts of 27\% (58 out of 218) of the stories were revised after initial generations.
Reported edits included adjusting vocabulary not aligned with a conversational tone (P2, P6--8, and P15) and expressions not typically found in children's literature (P8--9, P13, and P16). P5 also regenerated illustrations when they did not align with the characters. 

\subsubsection{Created Stories}
The parents created stories on various topics and target behaviors. 
In the 218 stories parents created, we identified 18 target behavior categories and grouped them into six higher-level semantics: \textit{social norms}, \textit{self-care \& daily living}, \textit{social interaction \& exchanges}, \textit{safety}, \textit{emotion \& self-regulation}, and \textit{challenges and new experiences} (\autoref{tab:result:behaviors} in Appendix). Each participant created stories of 7.38 unique categories on average (\reportstats{2.00}{5}{P1, P3, P6--7}{11}{P12}).

Parents created the largest number of stories about \textbf{social norms} (76/218; 35\%), such as following rules in shared spaces (\eg{}, \textit{keeping quiet in the library}) or norms for interacting with others. 
Parents also frequently created stories guiding \textbf{self-care and daily living} skills (54/218; 25\%), such as personal hygiene and eating habits. This aligns with literature that reports autistic individuals' challenges in adaptive functioning and daily living skills~\cite{Goldie2021Gap, Yela2021ADLASD}.
Thirteen stories (14\%) contained guidance on \textbf{social interaction and exchanges}. Notably, half of the parents created 16 stories that encourage their children to \textbf{self-express} their needs, preferences, or feelings, in verbal communication instead of non-verbal actions.
Given that autistic children tend to stay within familiar routines~\cite{larson2006caregiving}, several parents created stories encouraging their child to engage in \textbf{new activities}, \eg{} assembling toy bricks. 

Some stories addressed challenges associated with autism-specific traits. 18 stories included topics regarding \textbf{stimming}, self-stimulatory behaviors---such as wiggling fingers---that autistic individuals commonly demonstrate~\cite{Kapp2019Stimming, Rajagopalan2013selfstimulatory, Eason1982SelfStimulatory}. Most of them explicitly guided regulating such behaviors in relation to its disruption of social norms (\cf{}, `Stimming in public').
Two parents created three stories guiding the regulation of socially inconsiderate behaviors---such as persistently talking about a particular topic with friends---that may stem from \textbf{fixated interests}~\cite{gunn2016teaching}.



In debriefing, we asked parents how they derived such target behaviors. Parents reported that, beyond reinforcing general desirable behaviors (\eg{}, getting along well with siblings--P9~[relationship]), they often tailored stories to their child's immediate experiences. These included specific past incidents (\eg{}, peeing in the bathwater--P12~[hygiene]) and anticipated future events (\eg{}, attending a wedding--P1~[interpersonal norms]). This indicates that parents used \sysname{} not only to convey general norms but also to address concrete, situational needs that arose in daily life.

\subsubsection{Personalization Through Children's Interests}
\begin{figure}[t]
    \includegraphics[width=\columnwidth]{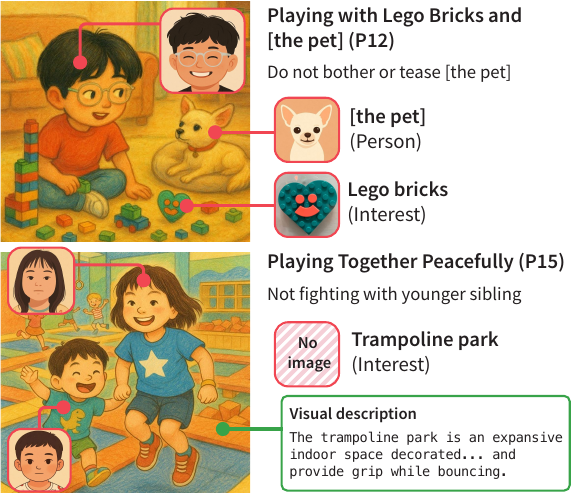}
    \caption{Illustrations from participants' stories and how contextual entities were reflected in them.}
    \label{fig:illustration_examples}
    \Description{Figure 8 presents excerpts of story illustrations that demonstrate how children’s interests, places, and persons chosen by parents were blended into personalized narratives.

Left panel: An illustration of children bouncing at a trampoline park. The input interest was “Trampoline,” which the system expanded into the setting of a Trampoline park. No photo was provided, but the description guided the generated image.
Center panel: An illustration of children playing with a toy train in a living room. This was generated using the child’s interest in “Train” and the place “Living room,” supported by photos provided by parents.
Right panel: An illustration of a boy playing with Lego bricks beside a pet dog. This scene was generated from the child’s interest in “Lego bricks” and the inclusion of [the pet] as a person entity, again guided by a parent-provided photo.

The bottom table shows the titles and target behaviors associated with these stories:
Left: Playing together peacefully (P15) → Target: Do not fight with younger sibling.
Center: Using Words instead of Hands (P7) → Target: Express refusal verbally instead of hitting.
Right: Playing with Lego Bricks and [the pet] (P12) → Target: Do not bother or tease the pet.

This figure illustrates how the system incorporated personal context items into both visuals and storylines, producing meaningful narratives tailored to each child.
}
\end{figure}

Across all created stories, we identified a total of 98 interests, which we grouped into ten categories (see \autoref{tab:result_interests} in Appendix), including \textbf{activity}, \textbf{character}, \textbf{object}, \textbf{sports}, \textbf{food}, \textbf{place}, \textbf{vehicle}, \textbf{person}, \textbf{animal}, and \textbf{other}. The most common category was activities (18\%)---such as Bubble play [C3]---followed by characters (15\%) and objects (15\%). On average, each dyad registered interests from 3.69 categories (\reportstats{1.96}{1}{P10}{8}{P14--15}). \autoref{fig:illustration_examples} curates some illustrations from two dyads and how their entities were reflected to the scenes.

Each parent used an average of 6.13 interests for creating books, with high variance across dyads (\reportstats{4.05}{3}{P6--7, P10}{17}{P14}). 
In debriefing, parents reported their own strategies of what and how to integrate interests to stories. For example, P4 explained that because C4's interests were narrow, she selected story elements that fit the target behavior and framed them in ways that would still engage C4, rather than relying only on their existing interests.


\subsubsection{Usability of \creator{}}

\begin{figure}[t]
    \centering
    \includegraphics[width=\columnwidth]{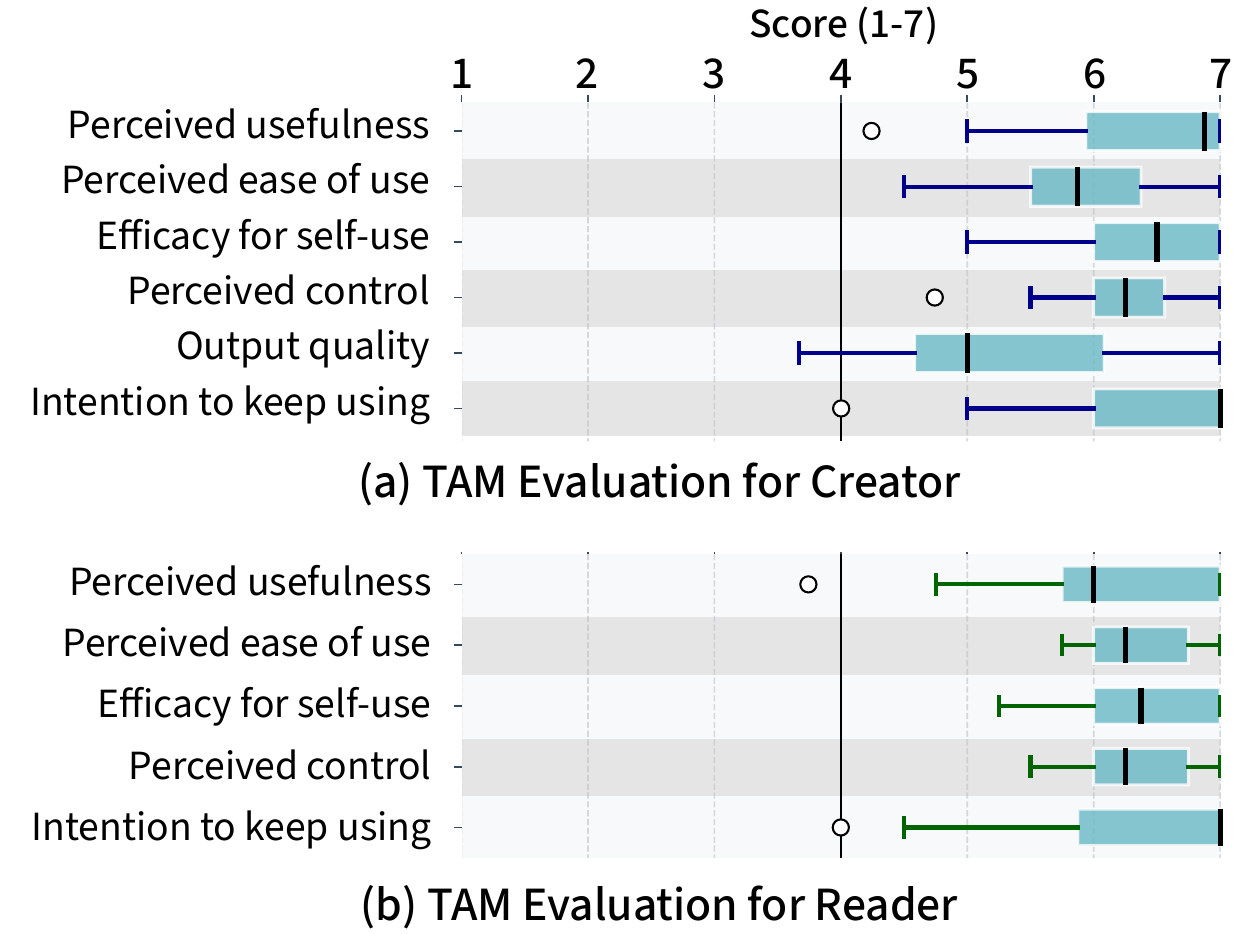}
    \caption{Post-study ratings of parent participants on the technology acceptance model (TAM) subscales, evaluating \creator{} (a) and \reader{} (b) separately. 
    }
    \label{fig:results_tam}
    \labelphantom{fig:results_tam:creator}
    \labelphantom{fig:results_tam:reader}
    \Description{Figure 7 presents boxplots of parent participants’ ratings on the Technology Acceptance Model (TAM) subscales for two applications: Creator (a, left) and Reader (b, right). Each subscale is rated on a 7-point scale (1–7).

Creator evaluation (a):
Perceived usefulness and intention to keep using received the highest median scores (close to 7).
Perceived ease of use, efficacy for self-use, and perceived control scored moderately high, with medians around 6.
Output quality received the lowest ratings, with a median near 5 and a wider spread of responses.

Reader evaluation (b):
Intention to keep using were consistently high, with medians around 7.
Perceived usefulness, perceived ease of use, ratings for efficacy for self-use, and perceived control also scored positively, with medians around 6–6.5.
The figure highlights differences in parent perceptions of the two apps, showing stronger acceptance for the Reader app.}
\end{figure}

Parents generally found the system intuitive and easy to use, as they rated the \textit{Perceived ease of use} of \creator{} with average scores of 5.94 on a scale of 1 to 7 (See \autoref{fig:results_tam:creator}).
In debriefing, parents with prior experience in creating Social Stories reported that \sysname{} was a lot more convenient than their previous methods. In particular, they appreciated faster story creation and the mobile web support, which allowed them to use spare moments to create content on smartphones (\eg{}, while waiting for their child during the therapy session).

At the same time, parents mentioned where the process could be further supported. While the story creation process was straightforward, some parents felt challenges in deciding which topics to address. P2 and P15 suggested the system could recommend new themes or automatically generate stories around suitable subjects. 

Overall, parents expressed satisfaction with the current story design and structure, while also suggesting directions for improvement. Five parents (P1, P3, P11-12, P15) suggested greater variety in story structures, as the narrative pattern sometimes felt monotonous and repetitive.
Some parents emphasized depicting more concrete consequences of children’s actions; P3 and P13 suggesting dramatic outcomes such as injuries for risky behaviors. 

\subsection{Story Reading}

In this part, we investigate how reading occurred, children’s reactions to personalized content, and their interactions with the branching story paths.

\subsubsection{Reading Activities}
Throughout the study period, participants read stories on a regular basis. On average, parents spent 5.21 minutes per day reading (\reportstats{3.67}{0.57}{D2}{16.8}{D15}) and completed 4.25 stories per day (\reportstats{1.75}{1.57}{D2}{7.14}{D15}). 

Parents observed that their children greatly enjoyed these reading activities and participated proactively. In daily surveys, parents rated the children’s engagement at an average of 4.02 on a 5-point Likert scale, indicating a high level of engagement. P9 shared, \textit{``On our scheduled reading time, my child brings the tablet and asks me to read the story.''} Overall, most parents reported that they did not need to employ additional strategies to sustain their child’s engagement.


\subsubsection{Children's Reaction to Personalized Content}

In debriefing, parents reported that incorporating their children's interests and contexts into the stories helped engage them in the reading activities. Many also noted that their children appeared more interested when encountering characters that resembles themselves or familiar people. 
Children often reacted positively when the story’s protagonist, modeled after themselves, engaged in desirable behaviors, smiling in response. Conversely, they showed reluctance when the undesirable path depicted embarrassing scenarios. 
Meanwhile, P10 observed her child overly fixated on specific interests, insisting on repeatedly viewing and discussing only those scenes. This tendency was also noted as a potential concern in expert interviews that personalization may also reinforce narrow focus. 

\subsubsection{Children's Interaction with the Story Paths}

Most parents found the multi-path story design helpful. They appreciated that the branching structure encouraged children to participate proactively (P7), pause and reflect (P10), experience the narrative of considering the other's emotions alongside their own (P6), and learn that there was no single correct answer (P2).

At the same time, several parents suggested the need for greater variety in the story paths, \eg{}, increasing the number of branching paths (P6) and providing more varied interactive features beyond multiple choices such as ``\textit{thought-provoking questions before showing the answer}'' (P11) and ``\textit{puzzles or connect-the-dots}'' (P15).

Some parents also observed that their child consistently chose only one type of path---either the desirable (P2, P4, P8, P13, P16) or the undesirable (P1, P10). This tendency was attributed to factors such as a compulsion to select the correct answer or the stimulating nature of the undesirable paths. In response, most parents encouraged their children to explore the alternative path, 
thereby helping them recognize the consequences of different choices.

\subsection{Parents' Reflections on Behavioral Guidance}
Based on debriefing comments and exit surveys, we summarize parents' post-study reflections on behavioral guidance.

\subsubsection{Reflections on Responding to Children's Behaviors}

Many parents reported that creating and reading stories with \sysname{} positively influenced their approach to addressing their child’s challenging behaviors. 
The exit survey indicates that in 82\% of the cases, their own responses to the behaviors became more positive (See \autoref{fig:result_behavior:parents} in Appendix for individual ratings). P8 remarked, \textit{``The process of creating the story became an opportunity to reflect on my child’s behavior and think more from their perspective.''} 


As children could naturally learn desirable behaviors while reading, many parents appreciated being able to address challenging behaviors through storytelling rather than scolding or nagging:  
``\textit{Normally, I would have nagged, but we could talk about what's right or wrong as part of the story, which really helped.} (P3)'' 
Parents reported that the stories helped them reconsider behaviors they had overlooked or just accepted because of autism: ``\textit{I started to look at my child with a longer-term perspective, realizing that behaviors I used to simply accept could, in fact, be changed.} (P9)''
This perspective shift extended to parents' overall mindset, focusing on what the child enjoys and does well rather than on what they cannot. P14 reflected, \textit{``Beyond the story, I made a commitment to talk together while engaging with the things my child likes and is good at.''}

\subsubsection{Reflections on Children's Behavior Changes}
\label{sec:result_change_child}
Parents reported positive perceived changes in 76\% of the target behaviors (\autoref{fig:result_behavior:children} in Appendix).
These were noted across various domains: trying a new ride at the playground (P5), overcoming fear of rain (P7), and getting along better with a younger sibling (P9). 
Parents referred to these outcomes---along with the ease of use of \sysname{}---as a key reason for their high ratings on the \textit{Intention to keep using} on the TAM survey (6+ for both \creator{} and \reader{}; 
See \autoref{fig:results_tam}). 
P7 reflected, \textit{``Honestly, I was skeptical about how much my child would change in just two weeks from reading this, but it turned out my child did adopt the target behavior. And since it wasn’t a burden for me, I want to continue using it.''} P9 also remarked, \textit{``My child, who used to be a picky eater, suddenly said they wanted to eat broccoli! I was so surprised at this change.''} 
%
While these observations are based on parents' self-reports from a short-term deployment, they still highlight \sysname{}'s potential as a practical tool for parents in guiding their children's social behaviors in real-life context.

\section{Discussion}

In this section, we further reflect on the design process of \sysname{} and its deployment, discussing lessons learned and implications for ethical and effective social narrative technologies for parents.

\subsection{Considerations on Parent-led Story Creation}

\sysname{} empowered parents to take an active role in creating social narratives, engage in the intervention, ground the stories in children’s real-life contexts by leveraging their knowledge of their children, and address both immediate challenges and long-term developmental goals. Parents also reported that the process of creating, reviewing, and reading stories prompted reflections on their parenting strategies and communication styles.

In the meantime, parent-driven topic selection also raises important considerations on appropriateness and safety. Parents may prioritize behaviors based on personal values or spontaneous frustrations rather than evidence-based intervention priorities~\cite{brookman2006parenting}, prone to misalignment with the child’s developmental needs. In our study, some parents attempted to modify behaviors closely tied to autism-specific traits (\eg{}, stimming), where overly controlling or punitive framing could impose emotional stress on the child~\cite{kapp2019people}. 

These risks highlight the need to balance the autonomy of parents with appropriate guidance and safeguards. For example, \sysname{} could incorporate scaffolding mechanisms or topic recommendations tailored to developmental levels~\cite{kim2013multisite}.
The system could also provide model stories that demonstrate supportive ways to address sensitive behaviors in supportive and constructive ways, enabling parents to create content that is both safe and therapeutically sound. In this way, parent-driven story creation can retain its advantages of personal relevance and practical applicability, while minimizing the risk of misdirection and more responsibly supporting the growth of autistic children.

\subsection{Implications for AI-based Behavioral Guidance Support}

Our study highlights important implications for leveraging GenAIs to support parents in guiding their autistic children through social narratives.
First, GenAI should function as a \textit{reflective aid} that supports parents' sensemaking, rather than simply producing content. Parents reported that the story authoring process encouraged them to reflect not only on their child’s behaviors but also on their communication strategies. 
Second, social narratives can serve as a medium for constructive, non-disciplinary guidance. By framing situations in calm and supportive language, systems can help parents move toward supportive interaction styles.
Third, personalized narratives can foster parent--child interaction. When integrated into everyday contexts, stories become shared reference points that enable ongoing dialogue.
For example, P16 reflected, \textit{``My child’s expressive language is implicit and limited, so I often struggled with what topics to bring up. The story became a good conversation starter, enabling questions and responses.''} 

In summary, our work suggests that AI systems for behavioral guidance should take into consideration the everyday contexts of parenting---how they approach, communicate, and engage with their children.



\subsection{Ethics \& Privacy with Child's Personal Data}


In designing \sysname{}, we incorporated several safeguards to mitigate ethical and privacy risks associated with using children's data, which prior work identifies as highly sensitive~\cite{bailey2021perspective, jiao2025llms, caetano2025neglected}.
At the same time, achieving fully local and end-to-end control over such data remains open design challenges. One possible approach is to use browser-based frameworks such as MediaPipe Face Landmarker~\cite{mediapipe} and ONNX Runtime~\cite{onnxruntime} enable on-device processing without transmitting raw images to external servers. By leveraging such approaches, systems could extract personally identifiable features locally and generate stylized or abstracted representations that preserve recognizable characteristics while avoiding direct exposure of identifiable data.

\subsection{Limitations and Future Work}

Our child participants in the deployment study were mostly boys, limiting our ability to examine potential gender differences in engagement and behavioral patterns. 
Nonetheless, this gender imbalance may be partly consistent with the male-to-female ratio of autism population in Korea (approx. 4:1) for children aged 7--12~\cite{kosis2025gender}.

Our work relied primarily on parent self-reports and system logs, consistent with \sysname{}'s design as a parent-facing tool. While parents provided valuable reflections and expressed willingness to use the system, incorporating direct evidence of children’s in-situ interactions could further strengthen the evaluation of \sysname{}’s practical potential. Although collecting such data from autistic children can be challenging, future research could incorporate complementary methods~\cite{spiel2019agency, wilson2019co, wilson2020self}---such as direct observation, video recordings, or child-centered measures of engagement---to provide a more holistic understanding of how children themselves experience and respond to narrative-based interventions.

\section{Conclusion}

We presented \sysname{}, a GenAI-based
system that supports parents to create personalized, multi-path social narratives and read them with their autistic children. Drawing from formative interviews with experts, \sysname{} is designed to reflect the children's personal interests and target behavior, and incorporate multi-path story structure that allows the children to rehearse behavioral options. Through a two-week deployment study with 16 parent--child dyads, we demonstrated how \sysname{} supports parents in the story creation process, engages children in reading activities, and shapes parental perspectives towards behavioral guidance. We discussed lessons learned from the study highlighting future research areas, including balancing parental autonomy and supporting ethical and privacy-considerate personalization.


\begin{acks}
We thank the expert participants of our formative interviews and the parents and children of the deployment study, for their time and efforts. We are also grateful to Migyeong Yang and Suwon Yoon for their feedback on our paper draft. This work was supported through a research internship at NAVER AI Lab of NAVER Cloud.
\end{acks}

\bibliographystyle{ACM-Reference-Format}
\bibliography{references/autism, references/storytelling, references/tools, references/etc}


\clearpage
\onecolumn
\appendix

\section{Formative Study}
\begin{table}[htbp]
\sffamily
\small
	\def\arraystretch{1.2}\setlength{\tabcolsep}{0.15em}
		    \centering

\caption{Demographic information of experts of the formative interviews.}
\label{tab:formative_demographics}
\Description{Table 1 provides demographic and professional background information of 10 experts (E1–E10) who participated in formative interviews.
Job titles: The experts represent a range of professions, including licensed counselor, K–12 teachers (elementary and special education), clinical psychologists, art education specialist, speech-language pathologist, ABA therapist, and child development specialist.
Years of experience: Participants had between 4 and 28 years of professional experience.
Age and gender: Ages ranged from 28 to 55 years; six were female and four were male.
Parenting experience: Four experts (E1, E2, E3, E10) reported raising a child with a developmental disability, while six did not.
This table highlights the diversity of professional expertise and lived experience among the interview participants.}
\begin{tabular}{|l!{\color{gray}\vrule}p{0.20\textwidth}!{\color{lightgray}\vrule}>{\raggedleft\arraybackslash}p{0.03\textwidth}l!{\color{lightgray}\vrule}c|}
\hline
\rowcolor{tableheader}
\textbf{Expert} & 
\textbf{Job title} & 
\multicolumn{2}{m{0.08\textwidth}!{\color{lightgray}\vrule}}{\textbf{Experience}} & 
\textbf{Child with Developmental Disabilities} \\
\hline
\arrayrulecolor{tablegrayline}
\textbf{E1} (44F) & Licensed counselor & 15 & years & Yes \\
\hline
\textbf{E2} (40F) & K-12 teacher (elementary) & 17 & years & Yes \\
\hline
\textbf{E3} (47M) & Clinical psychologist & 5 & years & Yes \\
\hline
\textbf{E4} (47F) & Clinical psychologist & 25 & years & No \\
\hline
\textbf{E5} (52F) & Art education specialist & 14 & years & No \\
\hline
\textbf{E6} (55M) & K-12 teacher (special education) & 28 & years & No \\
\hline
\textbf{E7} (28F) & K-12 teacher (special education) & 4 & years  & No \\
\hline
\textbf{E8} (45F) & Speech-language pathologist & 10 & years & No \\
\hline
\textbf{E9} (41M) & ABA therapist & 17 & years & No \\
\hline
\textbf{E10} (53M) & Child development specialist & 10 & years & Yes \\
\arrayrulecolor{black}\hline
\end{tabular}
\end{table}

\newpage

\section{Story Text Generation Prompt Templates}
\label{sec:appendix_story_generation_prompt}

\begin{figure}[h!]
    \centering
    \includegraphics[width=0.95\textwidth]{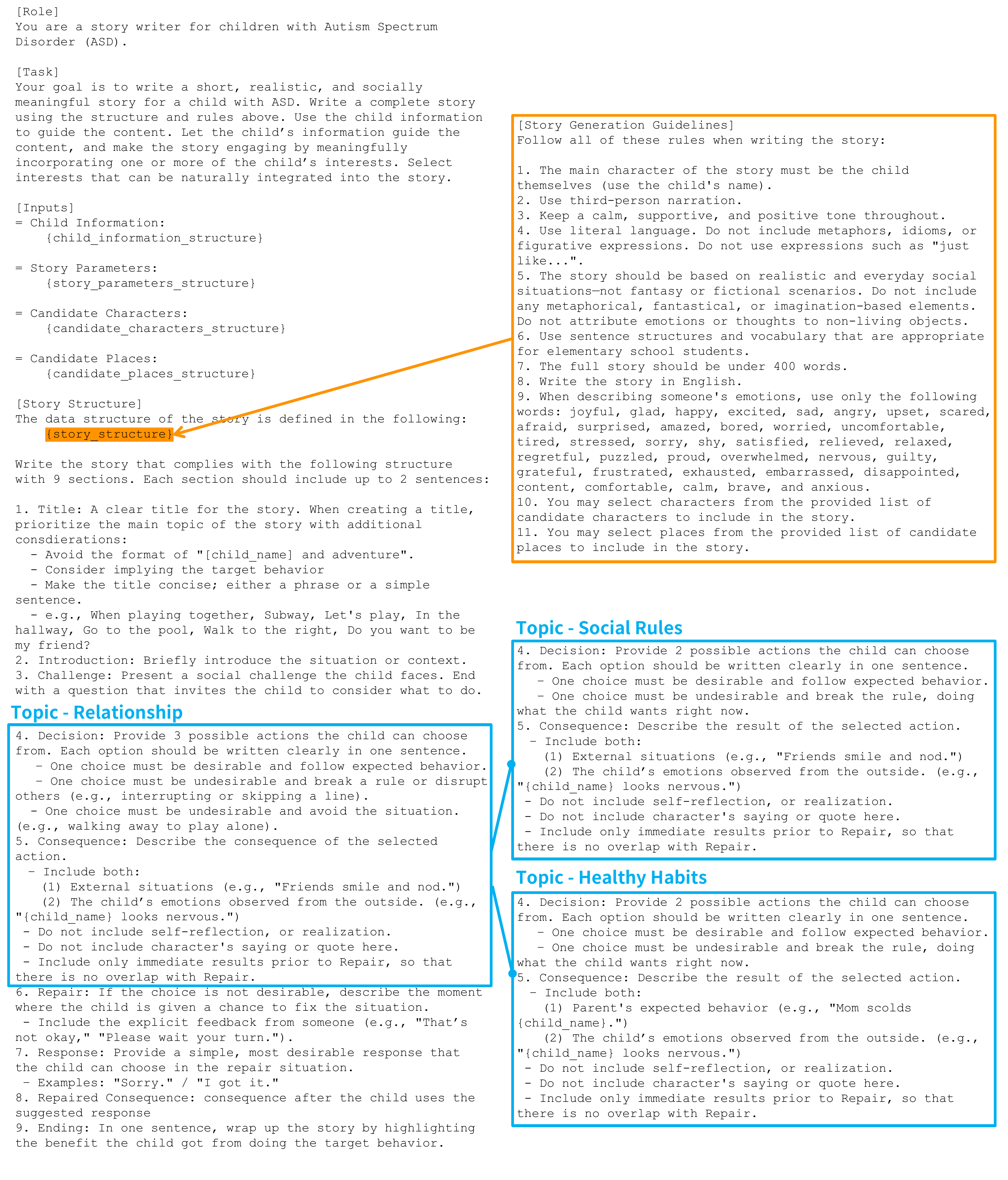}
    \caption{Prompt template for generating story text per topic.}
    \label{fig:story_generation_prompt}
\end{figure}

\clearpage
\section{User Interface}
\begin{figure}[h!]
    \centering
    \includegraphics[width=0.9\textwidth]{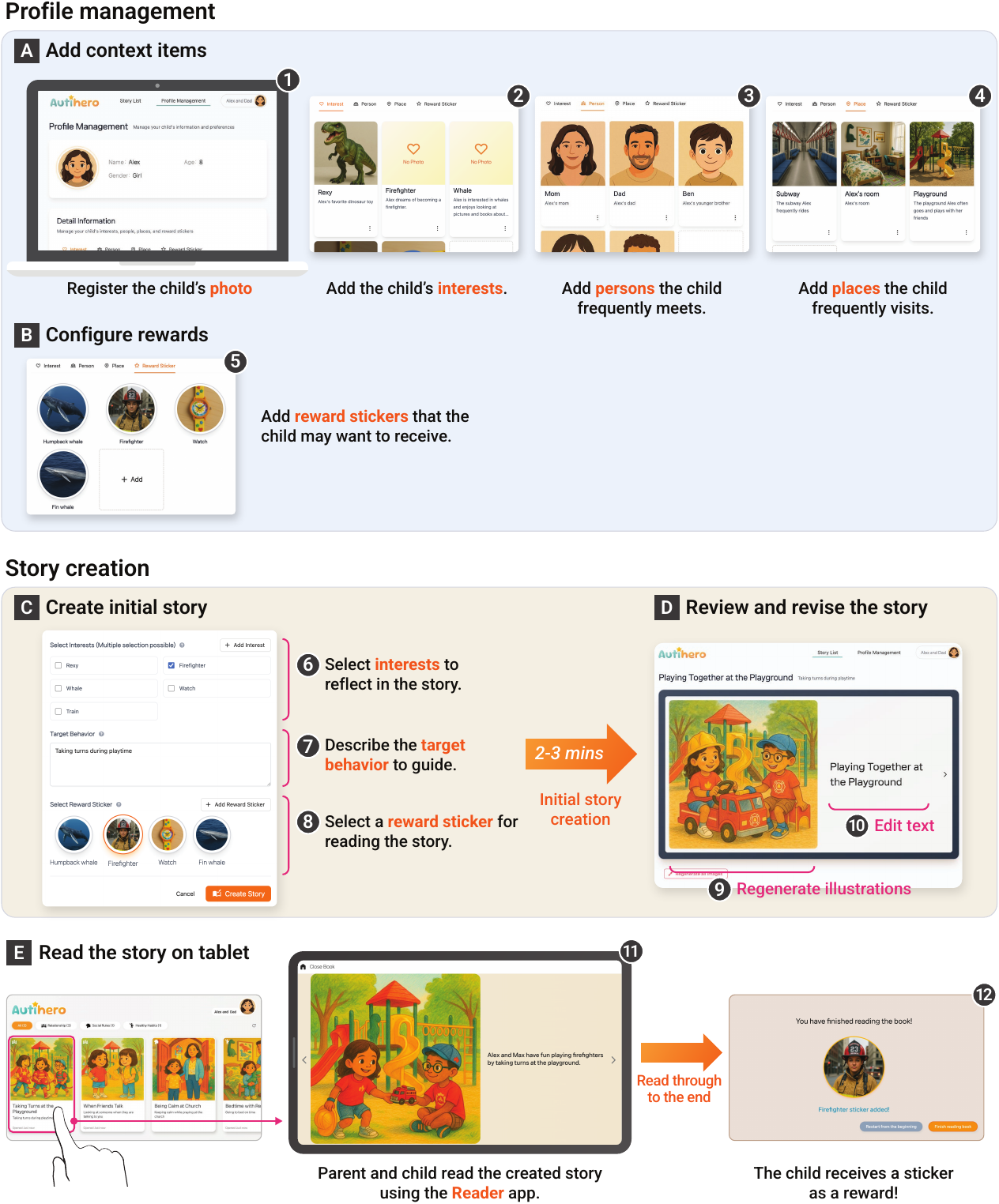}
    \caption{Overview of the usage flow of \sysname{}, from profile management to story creation, and story reading.}
    \label{fig:system_usageflow}
    \Description{Figure 3 illustrates the system usage flow, which consists of three main stages: Profile Management, Story Creation, and Story Reading.
Profile Management (steps 1–5): The parent uploads the child’s photo, adds interests, frequently met persons, and commonly visited places. Reward stickers that the child may want to receive are also added.
Story Creation (steps 6–10): The parent selects interests to reflect in the story, describes the target behavior for the child, and chooses a reward sticker. The system generates a complete story with both text and illustrations within 2–3 minutes. Parents can regenerate images to fix awkward illustrations or directly edit the text.
Story Reading (steps 11–12): The parent and child read the created story together using the Reader. After completing the story, the child receives the selected sticker as a reward.
The figure uses icons, screenshots, and arrows to visually represent each step in sequence, emphasizing the progression from personalization to reading and reward.
}
\end{figure}

\clearpage
\section{Deployment Study}
\begin{table*}[htbp]
\sffamily
\small
	\def\arraystretch{1.2}\setlength{\tabcolsep}{0.2em}
		    \centering
            
\caption{Demographic information of deployment study participants with children's CDC autism level and literacy description provided by parents. The primary/secondary labels of parent type indicate whether the parents self-identify themselves as primary or secondary caregivers. *Four children were assessed as having Level 2 autism in verbal communication but were considered to have potentially sufficient comprehension of images and text to participate in our study.}
\label{tab:deployment:demographic}
\Description{Table 2 summarizes demographic details of 16 families (P1–P16) who participated in the deployment study, including information about their autistic children (C1–C16).
Parent participants: Most were mothers identifying as primary caregivers. Two fathers were included, one of whom is primary caregiver and one of whom is a secondary caregiver.
Children participants: All children were autistic diagnosed with Level 1 or 2, aged 7–12 years; 14 were boys and 2 were girls.
Siblings: Several children had siblings, including brothers and sisters. Four children (C3, C6, C13, C14) were noted with “Level 2,” meaning Level 2 autism in verbal communication but sufficient comprehension for participation.
Literacy of short stories:
Some children enjoyed independent reading but preferred stories with illustrations or familiar topics (e.g., C4, C7, C12).
Others struggled with text-only books, abstract language, inference, or comprehension of longer passages (e.g., C8, C10, C11).
This table highlights the diverse reading abilities, preferences, and challenges of participating children, providing important context for evaluating AutiHero’s impact.}
\begin{tabular}{|cm{0.09\textwidth}!{\color{gray}\vrule}cc!{\color{tablegrayline}\vrule}c!{\color{tablegrayline}\vrule}c!{\color{tablegrayline}\vrule}c!{\color{tablegrayline}\vrule}m{0.55\textwidth}|}
\hline
\rowcolor{lightgray}
\multicolumn{2}{|l!{\color{gray}\vrule}}{\textbf{Parents}} & \multicolumn{6}{l|}{\textbf{Autistic Children}} \\
\rowcolor{tableheader}
\textbf{Alias} & \textbf{Type} & \textbf{Alias} & \multicolumn{1}{l}{\textbf{Age}} & \textbf{Gender} & \textbf{Siblings} & \textbf{Level} & \textbf{Literacy of Short Stories} \\
\hline
\cellcolor{tableheader}\textbf{P1} & Mother\newline{}(primary) & \cellcolor{tableheader}\textbf{C1} & 8 & Boy & & 1 & Enjoys nonfiction (\eg{}, science, economics); finds story- and emotion-centered literature less engaging and harder to understand. \\
\arrayrulecolor{tablegrayline}\hline
\cellcolor{tableheader}\textbf{P2} & Mother\newline{}(primary) & \cellcolor{tableheader}\textbf{C2} & 9 & Boy & & 1 & Listens to audiobooks at bedtime, memorizes content,\newline{}and tends to fixate on interests. \\
\hline
\cellcolor{tableheader}\textbf{P3} & Father\newline{}(secondary) & \cellcolor{tableheader}\textbf{C3} & 9 & Boy & A brother & 2* & Enjoys picture books and can read words,\newline{}but struggles to use content for reciprocal communication. \\
\hline
\cellcolor{tableheader}\textbf{P4} & Mother\newline{}(primary) & \cellcolor{tableheader}\textbf{C4} & 9 & Boy & & 1 & Reads and understands books better when pictures accompany text;\newline{}struggles with text-only books. \\
\hline
\cellcolor{tableheader}\textbf{P5} & Mother\newline{}(primary) & \cellcolor{tableheader}\textbf{C5} & 8 & Boy & & 1 & Prefers Baek Hee-na's books (\eg{}, Magic Candies) \\
\hline
\cellcolor{tableheader}\textbf{P6} & Mother\newline{}(primary) & \cellcolor{tableheader}\textbf{C6} & 9 & Boy & & 2*  & Can read independently and understand simple content;\newline{}recalls explicit details but struggles with inference. \\
\hline
\cellcolor{tableheader}\textbf{P7} & Mother\newline{}(primary) & \cellcolor{tableheader}\textbf{C7} & 8 & Boy & & 1  & Reads independently but prefers when the mother reads aloud. Understands better with illustrations and shows stronger interest in personally relevant topics. \\\hline
\cellcolor{tableheader}\textbf{P8} & Mother\newline{}(primary) & \cellcolor{tableheader}\textbf{C8} & 12 & Boy & A brother & 1 & Has difficulty understanding long text longer than one page; can answer factual questions about short texts but struggles with abstract or implied meaning. \\\hline
\cellcolor{tableheader}\textbf{P9} & Mother\newline{}(primary) & \cellcolor{tableheader}\textbf{C9} & 9 & Boy & A brother & 1 & Enjoys recalling storylines and characters; can answer general questions\newline{}but struggles with questions requiring interpretation of intent. \\\hline
\cellcolor{tableheader}\textbf{P10} & Mother\newline{}(primary) & \cellcolor{tableheader}\textbf{C10} & 8 & Girl & & 1 & Fairy tales and classics may be overstimulating; prefers realistic moral or leadership stories. Requires guided reading to understand character emotions and implied meanings. Finds books without illustrations difficult; struggles with metaphor, symbolism, and emotional nuance. \\\hline
\cellcolor{tableheader}\textbf{P11} & Mother\newline{}(primary) & \cellcolor{tableheader}\textbf{C11} & 7 & Boy &  & 1 & Understands simple passages but struggles with long or complex language. Reads kindergarten-level books with comprehension; can read knowledge-based books but has difficulty understanding them due to cognitive demands. \\\hline
\cellcolor{tableheader}\textbf{P12} & Mother\newline{}(primary) & \cellcolor{tableheader}\textbf{C12} & 7 & Boy &  & 1 & Enjoys independent reading, particularly with pictures to aid comprehension. \\\hline
\cellcolor{tableheader}\textbf{P13} & Mother\newline{}(primary) & \cellcolor{tableheader}\textbf{C13} & 8 & Boy & A sister & 2* & Understands short phrases or sentences, but comprehension decreases with longer or abstract text. Answers to content-related questions are limited, even when text is understood. \\\hline
\cellcolor{tableheader}\textbf{P14} & Mother\newline{}(primary) & \cellcolor{tableheader}\textbf{C14} & 8 & Boy & A sister & 2* & Prefers joint reading with mother or sibling; repeats favorite character content; struggles with overall story comprehension. \\\hline
\cellcolor{tableheader}\textbf{P15} & Father\newline{}(primary) & \cellcolor{tableheader}\textbf{C15} & 8 & Girl & A brother & 1 & Reads simple sentences but struggles with abstract concepts, long sentences, and ``when/why/how'' questions. \\\hline

\cellcolor{tableheader}\textbf{P16} & Mother\newline{}(primary) & \cellcolor{tableheader}\textbf{C16} & 10 & Boy & A sister & 1 & Sensitive to audio (autism trait); dislikes audio-only stories but enjoys visual story media (e.g., YouTube Pinkfong stories). Remembers stories well after repeated reading; previously read 1–3 storybooks per week.\\
\arrayrulecolor{black}\hline
\end{tabular}
\end{table*}

\newcommand{\categoryparbox}[1]{\parbox{0.11\textwidth}{#1}}

\begin{table*}[htbp]
\sffamily
\small
	\def\arraystretch{1.25}\setlength{\tabcolsep}{0.2em}
		    \centering
            
\caption{Categorization of target behaviors parents entered to create stories, number of stories and dyads, and example target behaviors and story titles created from them.}
\label{tab:result:behaviors}
\Description{Table 3 categorizes the target behaviors that parents entered to create stories, along with the number of stories, participating dyads, and example story titles.
Social norms (76 stories):
Shared space (33 stories): Following rules at home, school, or community (e.g., “Keeping quiet in the library”).
Interpersonal norms (17): Rules for interacting with others (e.g., “Not interrupting when others are talking”).
Stimming in public (14): Managing stimming behaviors in public (e.g., “Clapping at the Right Time”).
Pragmatic language (8): Using polite/context-appropriate words.
Respect for life (4): Acting ethically toward animals and living beings.
Self-care & daily living (54 stories):
Hygiene (19): Practicing personal cleanliness (e.g., “Washing hands before pizza”).
Eating habit (13): Healthy routines during meals.
Screen time control (11): Balancing digital media use.
Daily living skill (11): Independent self-care or household tasks (e.g., “Doing homework independently”).
Social interaction & exchanges (30 stories):
Self-expression (16): Expressing needs and feelings (e.g., “Is It Okay to Touch?”).
Relationship (14): Building and maintaining bonds (e.g., “Not fighting with younger brother”).
Safety (26 stories):
Safety for self (23): Preventing harm to oneself (e.g., “Crossing the street after checking the green light”).
Safety for others (3): Preventing harm to others (e.g., “Playing safely on the slide”).
Emotion & self-regulation (24 stories):
Emotion regulation (17): Managing emotional responses (e.g., “Calming down with Lulu”).
Stimming (4): Regulating self-stimulatory behaviors.
Fixated interests (3): Reducing excessive focus on a single topic (e.g., “Choosing the Right Story”).
Challenges & new experiences (8 stories):
Engaging in activities (4): Encouraging persistence in rewarding activities (e.g., “Completing a Lego project”).
Trying new activities (4): Trying novel or avoided tasks (e.g., “Exploring the Ocean Waves”).
Overall, the table shows that most stories focused on social norms and self-care/daily living skills, while fewer addressed challenges and new experiences.}
\begin{tabular}{|m{0.11\textwidth}!{\color{tablegrayline}\vrule}m{0.15\textwidth}m{0.2\textwidth}!{\color{gray}\vrule}rr!{\color{tablegrayline}\vrule}c!{\color{gray}\vrule}m{0.38\textwidth}|}
\hline
\rowcolor{tableheader}
\textbf{Semantics}                                       & \textbf{Categories} & \textbf{Definition}                                        & \multicolumn{2}{l!{\color{tablegrayline}\vrule}}{\textbf{Count}} & \textbf{Dyads}                                          & \textbf{Example Target Behaviors / (Story Titles)}                                                \\
\hline
\multirow{5}{*}{\categoryparbox{\textbf{Social\newline{}norms} (76)}}                        & \textbf{Shared space}   & Following rules at home,\newline{}school, and in community\newline{}settings             & 33          & 15\%          & 13                              & Keeping quiet in the library                                     \newline{}(Quiet Time at the Library) {[}D4{]}      \\
\arrayrulecolor{tablegrayline}\cline{2-7}
                                                     & \textbf{Interpersonal\newline{}norms} & Following rules for\newline{}interacting with people                          & 17          & 8\%           & 10                               & Not interrupting when others are talking                         \newline{} (Waiting to Speak) {[}D10{]}              \\
\arrayrulecolor{tablegrayline}\cline{2-7}
                                                     & \textbf{Stimming in\newline{}public}      & Managing stimming in\newline{}situations where it may\newline{}disturb others        & 14          & 6\%           & 7                                 & Clapping at appropriate times, such as celebrations              \newline{}(Clapping at the Right Time) {[}D8{]}     \\
\arrayrulecolor{tablegrayline}\cline{2-7}
                                                     & \textbf{Pragmatic\newline{}language}      & Using polite and context-\newline{}appropriate words                          & 8           & 4\%           & 5                                & Speaking politely to adults  \newline{} (Speaking Politely to Adults) {[}D4{]}    \\
\arrayrulecolor{tablegrayline}\cline{2-7}
                                                     & \textbf{Respect for life}       & Acting ethically toward\newline{}animals and living beings                                & 4           & 2\%           & 4                                & Not killing ants in the park \newline{}(Respecting Nature in the Park) {[}D9{]}  \\

\arrayrulecolor{darkgray}\hline
                                                     
\multirow{4}{*}{\categoryparbox{\textbf{Self-care \&\newline{}daily living}\newline{}(54)}}          & \textbf{Hygiene}                     & Practicing personal care\newline{}for cleanliness and health& 19          & 9\%           & 9                                                 & Washing hands upon arriving home \newline{} (Wash Hands Before Pizza) {[}D11{]}       \\
\arrayrulecolor{tablegrayline}\cline{2-7}
                                                     & \textbf{Eating habit}                & Following healthy routines\newline{}and manners during meals                 & 13          & 6\%           & 9                                & Eating side dishes without picky eating \newline{} (Trying New Foods with Friends) {[}D9{]}  \\
\arrayrulecolor{tablegrayline}\cline{2-7}                                                     
                                                     & \textbf{Screen time\newline{}control}         & Limiting and balancing\newline{}use of digital devices                       & 11          & 5\%           & 9                                & Watching mukbang YouTube videos in moderation \newline{} (Mukbang Time Management) {[}D8{]}        \\
\arrayrulecolor{tablegrayline}\cline{2-7}                                                     
                                                     & \textbf{Daily living skill}          & Performing basic self-care\newline{}and household tasks\newline{}independently       & 11          & 5\%           & 7                                 & Doing homework independently \newline{} (Homework Time) {[}D12{]}                 \\

\arrayrulecolor{darkgray}\hline

\multirow{2}{*}{\categoryparbox{\textbf{Social\newline{}interaction \&\newline{}exchanges}\newline{}(30)}} & \textbf{Self-expression}              & Expressing needs,\newline{}preferences, or feelings& 16          & 7\%           & 8                                                         & Asking ``Can I borrow it?'' when lending something from friend \newline{} (Is It Okay to Touch?) {[}D7{]}           \\
\arrayrulecolor{tablegrayline}\cline{2-7}
                                                     & \textbf{Relationship}                & Building and maintaining\newline{}social bonds                               & 14          & 6\%           & 9                                & Not fighting with younger brother \newline{} (Playing Together Peacefully) {[}D15{]}   \\
\arrayrulecolor{darkgray}\hline
                                                     
\multirow{2}{*}{\textbf{Safety} (26)}                              & \textbf{Safety for self}             & Acting to prevent harm\newline{}to oneself                          & 23          & 11\%          & 8                               & Crossing the street after checking the green light \newline{} (The Green Light Guide) {[}D13{]}         \\
\arrayrulecolor{tablegrayline}\cline{2-7}
                                                     & \textbf{Safety for others}           & Acting to prevent harm\newline{}to others                    & 3           & 1\%           & 2                                & Not pushing friends when going down the slide                    \newline{} (Playing Safely on the Slide) {[}D14{]}   \\
\arrayrulecolor{darkgray}\hline

\multirow{3}{*}{\categoryparbox{\textbf{Emotion \&\newline{}self-regulation}\newline{}(24)}}          & \textbf{Emotion\newline{}regulation}        & Recognizing and\newline{}controlling one’s\newline{}emotional responses                & 17          & 8\%           & 9                                & Calming oneself quickly when feeling upset \newline{} (Calming Down with Lulu) {[}D1{]}         \\
\arrayrulecolor{tablegrayline}\cline{2-7}
                                                     & \textbf{Stimming}                    & Regulating self-\newline{}stimulatory behaviors                              & 4           & 2\%           & 2                                 & Not wiggling fingers while reading books \newline{} (Studying Calmly with Pinkfong) {[}D15{]} \\
\arrayrulecolor{tablegrayline}\cline{2-7}                                                     
                                                     & \textbf{Fixated interests}           & Regulating excessive focus\newline{}on a specific object,\newline{}activity, or topic & 3           & 1\%           & 2                                & Not telling monster stories that friends dislike \newline{} (Choosing the Right Story) {[}D12{]}      \\

\arrayrulecolor{darkgray}\hline
\multirow{2}{*}{\categoryparbox{\textbf{Challenges\newline{}and new\newline{}experiences} (8)}}                              & \textbf{Engaging in\newline{}activities}             & Encouraging effortful yet\newline{}rewarding activities                          & 4          & 2\%          & 3                              & Completing a lego project \newline{} (Building Together) {[}D2{]}         \\
\arrayrulecolor{tablegrayline}\cline{2-7}
                                                     & \textbf{Trying new\newline{}activities}           & Encouraging new or\newline{}previously-avoided\newline{}activities                     & 4           & 2\%           & 2                                 & Trying underwater diving \newline{} (Exploring the Ocean Waves) {[}D5{]} \\     
\arrayrulecolor{black}\hline

\end{tabular}
\end{table*}

\begin{table*}[htbp]
\sffamily
\small
	\def\arraystretch{1.25}\setlength{\tabcolsep}{0.2em}
		    \centering

\caption{Categories of children’s interests reflected in created stories, with representative examples and frequency of appearance.}
\label{tab:result_interests}
\Description{Table 4 summarizes the categories of children’s interests that were incorporated into the created stories, with representative examples and frequency counts.
Activity (18, 18\%) — e.g., bubble play, hotel trip, drawing, hanging, traditional drum (janggu) playing.
Fictional/symbolic characters (15, 15\%) — e.g., Poli, Korean traditional wedding bride, Pororo, TiniPing, Pinkfong.
Object (15, 15\%) — e.g., toy car, Lego, comic book, AI robot, circle (object).
Sports (11, 11\%) — e.g., soccer, swimming, running, inline skating.
Food (10, 10\%) — e.g., tteokbokki (spicy rice cakes), pork cutlet, kimchi stew, cucumber, hamburger.
Place (8, 8\%) — e.g., judo gym, water park, haunted house, escalator, fish café.
Vehicle (8, 8\%) — e.g., kickboard (scooter), train, bicycle, subway.
Person (7, 7\%) — e.g., friends, younger brother, younger sister.
Animal (5, 5\%) — e.g., pet dog, cat, stingray, shark.
Other (1, 1\%) — e.g., disaster scenarios like earthquake or tornado.
The table highlights the wide variety of personal and cultural interests represented in children’s stories, with activities and fictional characters being the most frequently included.}
\begin{tabular}{|m{0.15\textwidth}|m{0.75\textwidth}!{\color{tablegrayline}\vrule}r|}
\hline
\rowcolor{tableheader}
\textbf{Category} & \textbf{Examples} & \textbf{Count} \\
\hline
Activity & Bubble play [C3], Hotel trip [C8], Drawing [C12], Hanging [C14], \newline{}Janggu (Korean drum) playing [C15] & 18 (18\%) \\
\arrayrulecolor{tablegrayline}\hline
Fictional/symbolic characters & Poli [C1], Korean Traditional wedding bride [C10], Pororo [C11], TiniPing [C14], Pinkfong [C15] & 15 (15\%) \\
\hline
Object & Toy car [C1], Lego [C2], Comic book [C3], AI robot [C7], Circle (object) [C14] & 15 (15\%) \\
\hline
Sports & Soccer [C6, C9], Swimming [C4, C12, C13, C14], Running [C14], Inline skating [C14, C15] & 11 (11\%) \\
\hline
Food & Tteokbokki (spicy rice cakes) [C3], Pork cutlet [C11], Kimchi stew [C14], Cucumber [C15],\newline{}Hamburger [C16] & 10 (10\%) \\
\hline
Place & Judo gym [C2], Water park [C2, C5], Haunted house [C6], Escalator [C14], Fish cafe [C15] & 8 (8\%) \\
\hline
Vehicle & Kickboard (scooter) [C2], Train [C3, C7, C9], Bicycle [C4, C5, C14], Subway [C7] & 8 (8\%) \\
\hline
Person & Friend [C4, C8, C9, C14], Younger brother [C9, C15], Younger sister [C14] & 7 (7\%) \\
\hline
Animal & Pet dog [C1, C15], Cat [C8], Stingray [C15], Pet cat [C15], Shark [C15] & 5 (5\%) \\
\hline
Other & Disaster (\eg{}, earthquake, tornado) [C5] & 1 (1\%) \\
\arrayrulecolor{black}\hline
\end{tabular}
\end{table*}

\begin{figure*}[h!]
    \centering
    \includegraphics[width=\textwidth]{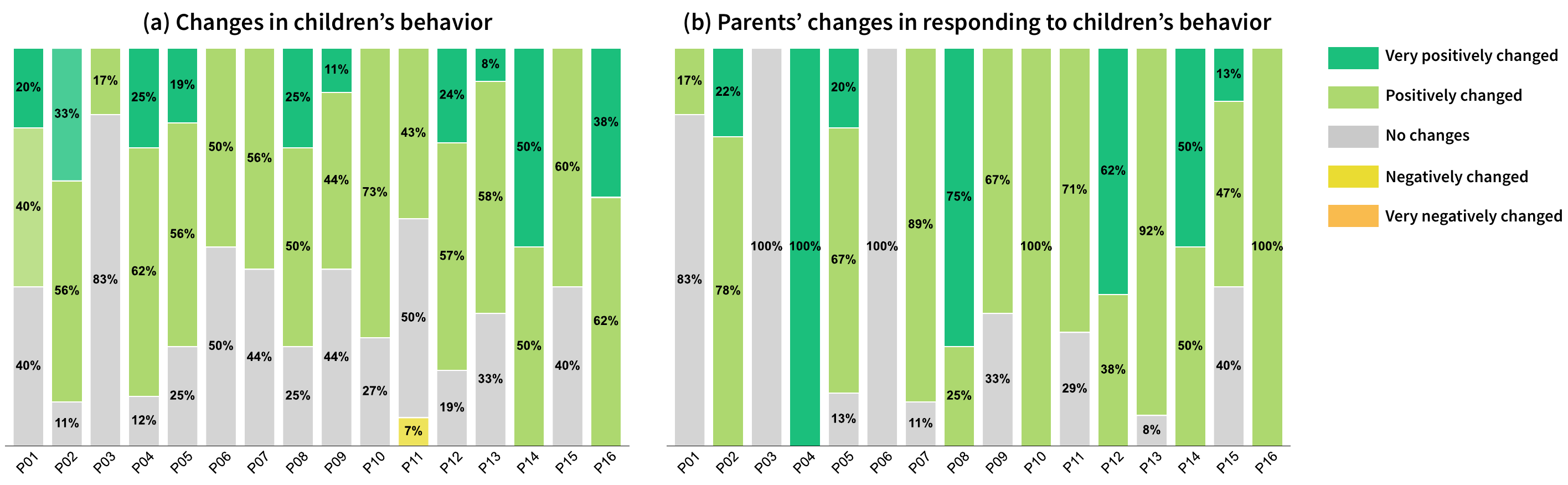}
    \labelphantom{fig:result_behavior:children}
    \labelphantom{fig:result_behavior:parents}  
    \caption{Reported changes in children’s behaviors (a) and parental behaviors (b) in response to children’s behaviors, based on the exit surveys with parents. The y-axis denotes the ratio of target behaviors against total number of behaviors per participant.}
    \label{fig:result_behavior}
    \Description{Figure 10 shows stacked bar charts summarizing reported changes in children’s behaviors (left) and parental responses to children’s behaviors (right), based on debriefing sessions with 16 parents (P1--16).

Color coding:
Blue = Very positively changed
Green = Positively changed
Gray = No change
Yellow = Negatively changed
Purple = Very negatively changed

Children’s behaviors (left):
Most participants reported positive or very positive changes.
One participant (P3) reported little change, with P11 noting a small negative effect (7

Parental responses (right):
Almost all parents reported positive or very positive changes in how they responded to their children’s behaviors.
For several parents (P4, P8, P10, P12, P14, P16), responses were entirely positive (100
No parents reported negative or very negative changes in their own behaviors.

The figure highlights that parents observed more consistent positive change in their own responses compared to children’s behaviors, suggesting stronger immediate impact on parental adaptation.}
\end{figure*}

\end{document}